\documentclass[twocolumn, eqsecnum, aps,showpacs, epsfig]{revtex4}
\usepackage{graphics}
\usepackage{dcolumn}
\begin{document}
\title{A Discrete Variational Approach for Investigation of Stationary
Localized States In A  Discrete Nonlinear
Schr$\ddot{\rm {\bf {o}}}$dinger Equation, Named IN-DNLS}
\author{K. Kundu}
\affiliation{ Institute of Physics, Bhubaneswar 751005, Orissa,
 India.}
\date{\today}
\begin{abstract}
IN-DNLS, considered here is a countable infinite set of coupled one
dimensional
nonlinear
ordinary differential difference equations with a tunable nonlinearity
parameter, $\nu$. This equation is continuous in time and discrete in
space with lattice translational invariance and has global gauge
invariance. When $\nu = 0$, it reduces to the famous integrable Ablowitz 
 - Ladik (AL) equation. Otherwise it is nonintegrable. The formation of
unstaggered and staggered stationary localized states (SLS) in IN-DNLS
is studied here using discrete variational method. The appropriate
functional is derived and its equivalence to the effective Lagrangian is
established. From the physical consideration, the ansatz of SLS is assumed
to have the functional form of stationary soliton of AL equation. So, the
ansatz contains three optimizable parameters, defining
width ($\beta^{-1})$, maximum amplitude and its position ($\sqrt\Psi$,
$x_{0}$). Four possible situations are considered. An unstaggered SLS can
be either on-site peaked $(x_{0} = 0.0)$ or inter-site peaked $(x_{0} =
0.5)$. On the other hand, a staggered SLS can be either Sievers-Takeno (ST) 
like mode $(x_{0} = 0.0)$, or Page(P) like mode $(x_{0} = 0.5)$. It is
shown here that unstable SLS arises due to incomplete consideration of the
problem. In the exact calculation, there exists no unstable mode. The
width of an unstaggered SLS of either type decreases with increasing $\nu
> 0$. Furthermore, on-site peaked state is found to be energetically
stable. These results are explained using the effective mass picture.
For the staggered SLS, the existence of ST like mode and P like mode 
is shown to be a fundamental property of a system, described by IN-DNLS.
Their properties are also investigated. For large width
and small amplitude SLS, the known asymptotic result for the amplitude is
obtained. Further scope and possible extensions of this work are  
discussed.

\end{abstract}

\maketitle

\section{introduction}

The study of energy localization in nonlinear lattices has become an
important field of research in nonlinear dynamics in the past couple of
decades\cite{1}. In this context, the subject of intrinsic localized modes
(ILM) has drawn a considerable attention as it offers appealing insights 
into a variety of problems ranging from the nonexponential energy
relaxation\cite{2} in solids, to the local denaturation of DNA double
strands\cite{3}. The subject is also an intense field of study in
material science, and nonlinear optic applications\cite{4,5}.

The necessary condition for the formation of intrinsic localized modes 
(ILMs) or excitations in translationally invariant nonlinear systems is
the balance between nonlinearity and dispersion. Furthermore, by localized
it is meant that the amplitude of such modes goes to zero at the
boundaries of the system, which is taken to be infinitely large. In other
words, the relevant localization length scale is much much smaller than
the system size length scale. There are two broad classes of intrinsic
localizations in (1 + 1) dimensional nonlinear continuous
systems\cite{6}. Shape preserving localized excitations, arising  in
nonlinear continuous systems by satisfying the above mentioned balancing
condition are called dynamical solitons\cite{7,8,9}. Solitons in
continuous nonlinear Schr$\ddot{\rm{o}}$dinger equation (cNLS) is an
example of dynamical solitons\cite{9,10}. By
solitons we usually mean moving shape preserving nonlinear excitations,
though there can be stationary solitons also. Take for example cNLS. One 
particular one-soliton solution of this equation is a stationary
soliton\cite{10}. Breathers belong to the second category of ILM in
nonlinear systems\cite{6}. Breathers are
spatially localized time periodic solutions of nonlinear equations. They
are characterized by internal oscillations\cite{6, 7, 11,12,
13,14,15,16,17,18, 19,20, 21, 22}. Again, by breathers we usually imply
stationary localized excitations in nonlinear systems. However, under
appropriate conditions, nonlinear systems may have
moving breathers\cite{7}. As for examples, we note that breathers can be
found in continuous systems, described by sine-Gordon (sG) equation and
modified KdV (mKdV) equation\cite{7}. Even in cNLS, the stationary one-
soliton solution is nothing but a breather\cite{12}. So, the distinction
between solitons and breathers is not always very rigorous. Breathers are,
however rare objects in continuous nonlinear equations and are usually
unstable\cite{12}. It is important in the present context to  note that
continuous nonlinear equations may have Galilean or Lorentz
invariance. For example, KdV and cNLS are Galilean invariant\cite{7}. So,
a soliton of some fixed amplitude in the cNLS and KdV can be Galileo
boosted to any velocity. Similarly, sG equation has both stationary and
moving breather solutions\cite{7,22}. These two solutions are, however,
connected by Lorentz transformation\cite{7}. So, in dealing with
stationary ILMs, we consider that moving frame which is at rest with
respect to the ILM.  

However, models describing microscopic phenomenon in condensed matter
physics are inherently discrete, with the lattice spacing between atomic
sites being a fundamental physical parameter. For these systems, an
accurate microscopic description involves a set of coupled ordinary 
differential-difference equations (ODDE). Coupled ODDEs are also
encountered in the study of many important problems in optics and other
branches of science\cite{12,22,23}. So, it is pertinent to discuss next
what features of continuous nonlinear equations are possibly destroyed and
what novel features can arise from the discretization of at least one of
the variables, say one spatial dimension.   

In the general discrete case, Galilean or Lorentz invariance in relevant
dynamical equations may not be present at all or may not be transparent at
the equation level. Consider, for example the AL\cite{24,25}  and the N-AL
equations\cite{26}. First one is the example of an integrable
nonlinear differential discrete equation, which is often referred
to as the integrable discretization of the cNLS equation. 
The other equation provides an example of a differential discrete
nonintegrable nonlinear equation, having solitary wave solutions. Most
importantly, the existence of solitary waves in the N-AL equation can be
shown analytically\cite{26}. The solitary wave solutions of these
equations have continuous translational symmetry, which can be seen from
the  analytical expression of the one-soliton solution of the AL
equation.  This, in turn implies that both the AL and the N-AL equations
have the Galilean invariance. So, also in case of ODDEs, stationarity in
the ILM will imply the moving frame which is at rest with respect to the
ILM.

The replacement of the spatial derivatives by spatial differences
in the equation of motion implies the reduction of symmetry of the
Hamiltonian, for systems executing Hamiltonian dynamics. In general,
lowering the symmetry means enriching the class
of solutions, because less restrictions are imposed. Of course, solutions
are also lost by lowering the symmetry - namely ones which are generated
by higher symmetry\cite{12}. Let us consider in this context two discrete
nonlinear equations, the Frenkel-Kontorova (FK)\cite{22, 27} and the
discrete nonlinear Schr$\ddot{\rm{o}}$dinger(DNLS) equations\cite{28}. 
These are obtained by standard discretization of sG and cNLS 
respectively\cite{22, 23, 28, 29}. The FK model can be used to
describe a broad spectrum of nonlinear physically important phenomena,
such as propagation of charge-density waves, the dynamics of absorbed
layer of atoms on crystal surfaces, commensurable-incommensurable phase
transitions, domain walls in magnetically ordered  structures
etc\cite{22,27}. On the other hand, to name a few, DNLS has been  used to
model the self-trapping phenomenon in nonlinear waveguide arrays\cite{23},
to investigate a slow coherent transport of polarons in (1+1) dimension in
condensed matter physics\cite{28}, and to study the dynamical phase
diagram of dilute Bose-Einstein condensates\cite{6}. We note that both of
these discrete equations are nonintegrable while their continuous versions
are integrable. It is relevant in this context to know  that kink and
antikink solutions of sG equation, which is the continuous integrable
version of FK model are moving topological solitons, and they arise due to
the balance between nonlinearity and constraints originating from
topological invariants in the system\cite{9}. On the other hand, there
exists no steady-state solutions for a moving kink in the FK model. What
we obtain instead is static kinks\cite{22}. To understand this, we note
that the uniform discretization of space variables transforms continuous
translational invariance to lattice translational invariance. This, in
turn leads to a periodic arrangement of Peierls-Nabarro
(PN) potential\cite{22,30}. Therefore, while
the continuous translational invariance leads to zero frequency Goldstone
modes in the system, discreteness introduces the PN barrier, with the
barrier energy E$_{\rm{PN}}$\cite{22}. Due to this potential any moving
kink radiates phonons and loses energy (E$_{\rm{kink}}$).  When $
E_{\rm{kink}} < E_{\rm{PN}}$, the kink is trapped in one of the potential
wells and further loss of energy by the kink by radiation of phonons takes
it to the bottom of the well. This, in turn yields static
kinks. Similarly, sG breathers arise due to the high symmetry of the
equation, and consequently are unstable towards perturbation\cite{12}. As
the discreteness in space variables act as an external symmetry breaking
perturbation, even a weak discreteness does not allow oscillating breather
modes exist as dynamical eigenmodes of the sG chain, and breathers are
destroyed by radiation of linear waves. In case of DNLS, similar analysis
has been done in a perturbative frame using AL one-soliton solution as
the zeroth order approximation\cite{28,31}. This analysis also shows that
discreteness introduces a trapping potential for moving solitons and when
discreteness exceeds a critical value, solitonic modes are trapped leading
ultimately to pinned or stationary solitons.        

It is already mentioned that the AL equation is an integrable discrete
nonlinear equation. More specifically, the said  equation is a countably 
infinite set of one dimensional nonlinear ordinary differential
difference equations. This equation is continuous in time, but discrete 
in space with lattice translational invariance.  The exact one-soliton
solution of the AL equation is characterized by two parameters, namely,
$\beta\ \in [0, \infty)$ and $k\ \in\ [- \pi, \pi]$\cite{24,25}. For each
$\beta$, there exists a band of velocities,determined by the other
parameter, $k$, at which the soliton can travel without experiencing any
PN pinning from the lattice discreteness\cite{32}. Consider now other
nonlinear equations in this series, namely  the N-AL equation\cite{26},
the modified Salerno equation (MSE)\cite{28, 33} and the IN-DNLS\cite{32}. 
All these equations are nonintegrable extension of the AL equation,
containing tunable nonlinearities. The N-AL equation is postulated and
investigated to study the effect of dispersive imbalance on the
maintenance of the moving solitonic profile. The importance of this
equation lies in its appearance in the dynamics of vibrons and excitons in
soft molecular chains\cite{26, 28}. The solitary wave solutions of this
equation are also characterized by the same two AL parameters, $\beta$ and
$k$. However, only certain values of $k$ are allowed, though $\beta$ can
take all possible permissible values. At the allowed values of $k$, the
term which imparts nonintegrability disappears. This, in turn makes the
solitary waves transparent to the PN potential, arising from the lattice
discreteness. For other values of $k$, the initial AL one-soliton
profiles are observed numerically to leave phonon tails behind, causing
both slowing down and distortion of the initial profile. Important
too in this context is an analytical investigation in a perturbative
framework of the dynamics of a moving AL soliton, described by the
N-AL equation. This analysis suggests that any moving soliton having
energy below the PN barrier, induced by the discreteness in the lattice
will be pinned, yielding thereby stationary solitons\cite{26}.   

The IN-DNLS is a hybrid form of the AL equation and the DNLS, again with
a tunable nonlinearity, the tuning of which switches the equation from the
integrable AL equation to the nonintegrable DNLS\cite{32}. To gauge the
physical significance of this equation, we mention the followings. This
equation is studied to investigate the discreteness induced oscillatory
instabilities of dark solitons\cite{34}. Furthermore, a discrete
electrical lattice where the dynamics of modulated waves can be modeled by
this equation is studied to investigate the modulation instability of
plane waves\cite{35}. In the MSE, the usual DNLS is replaced by a modified
version of DNLS, the ADNLS, which involves acoustic phonons in stead of
optical phonons in condensed matter physics parlance\cite{28, 31}. The
study of this equation is also important in understanding the dynamics of
vibrons and excitons in soft molecular chains. It is important to note
that both IN-DNLS and MSE investigate the competition between the on-site
trapping and the solitonic motion of the AL soliton\cite{28, 32, 36}. So,
the dynamics of a moving self-localized pulse, like the AL soliton in the
framework of the IN-DNLS or the MSE will be subjected to two important
effects. First one is the PN pinning arising from the lattice discreteness
and second one is a nonlinear interaction potential, trying to trap or
detrap the localized pulse.The cumulative effect of these two interactions
is expected to be the collapse of the  moving self-localized states to
stable, but pinned solitons. This has indeed been observed in a numerical
simulation\cite{32}. From this discussion so far, it can be concluded that
the sufficient condition to see the effect of discreteness on the dynamics
of nonlinear excitations is that the discrete nonlinear equations must be
nonintegrable. This nonintegrability can arise directly from the
discretization of the continuous nonlinear equations or by adding
integrability breaking terms to integrable discrete nonlinear equations.

Two important linear PDEs, which play very important roles in physics in
linear systems are free-particle Schr$\ddot{\rm{o}}$dinger equation and
the wave equation respectively\cite{37}. These equations are, of course
used to describe dynamics in continuous systems. The eigenvalue spectra of
these equations are continuous function of a parameter, ${\bf{k}}$, called
wave vector, with the $\overline {lim}\ =\  \infty$ and the 
$\underline{lim}\ =\ 0$ . In case of systems, described by
Schr$\ddot{\rm{o}}$dinger equation with  a single-particle potential, an
attractive potential will create localized states below the spectra and
these are called 'bound states' of the system\cite{37}. Furthermore, in
one (1 + 1) dimensional systems, even an infinitesimally small attractive
potential will create an exponentially localized bound state. On the other
hand, wave equation being second order in time, even in (1+1) dimension no
attractive potential, however large can create bound states. On the
contrary, one can get resonances from attractive potentials.

When the continuity in spatial variables is replaced by lattice
continuity, the continuous spectra of linear PDEs fragments into
bands. The number of bands will depend on the number of lattice points
in the unit cell. When linear substitutional impurities are added to
systems, described by a discrete Schr$\ddot{\rm{o}}$dinger equation,
spatially localized states are formed in the gap between
bands\cite{37, 38}. We note that for a state to be localized and stable,
it must be in the gap of the spectra. Furthermore, these states being
exact eigenstates of the relevant Hamiltonian, are stationary localized
states (SLS). For finite number linear impurities in (1+1) dimension, it
can be shown that the number of spatially exponentially localized states
cannot exceed the number of impurities and there must be at least one
exponentially localized state\cite{37}. On the other hand, almost all
states are exponentially localized in fully disordered (1 + 1) dimensional
systems\cite{38}. However, with correlated disorder, it is possible to
have some delocalized states\cite{39, 40}. In stead of linear impurities,
if finite number of nonlinear impurities are present, we again obtain SLS
in such systems. This can be analytically shown in the systems, described
by the DNLS\cite{41,42}.

The spatially discrete analog of the continuous wave equation is the
coupled mass-spring systems, with springs obeying the Hooke's
law\cite{37,38}. Here again we get bands of eigenmodes, depending on the
number of mass-spring units in a unit cell. The lowest band is called
acoustic branch, which describes the collective motion of the
masses. Other bands give optical phonons\cite{43}. In systems containing
finite number of mass impurities, only  light mass impurities will form
exponentially localized states above the acoustic band in (1 +
1) dimension. Similar result is also obtained with impurity in
springs\cite{37}. Here also almost all states are exponentially
localized in totally disordered  systems, whether the disorder is in the
mass or in the spring or in both\cite{38}. However, no states are obtained
below the acoustic branch. Most importantly, states around zero frequency
remain delocalized\cite{38}. In this system also, one can have nonlinear
impurities, either in the spring, or in the on-site potential or in
both. Any such impurity will produce SLS in the system\cite{44}. We end
this discussion by noting that both continuous and  discrete linear
systems cannot sustain any localized mode without broken continuous and
lattice translational invariance respectively.

A uniform discrete nonlinear system will have lattice translational
invariance. Similar to continuous nonlinear systems with translational
invariance, nonlinearity in discrete systems can also generate localized
modes by balancing the delocalization effect without requiring broken
periodicity. Such localized self-organization are the ILMs of discrete
nonlinear systems. It is important to note that ILMs of a discrete
nonlinear system are the exact eigenmodes of the nonlinear Hamiltonian,
describing the system. As in continuous systems, ILMs in
discrete systems can also be divided in two broad categories,
solitons and breathers\cite{11,12, 13, 14, 15, 16, 17, 18, 19, 20, 21, 
22}. In this case also the separation line is not always
distinct. Consider for example the AL equation. The stationary one-soliton
solutions of this equation are nothing but breathers\cite{24, 25}. ILMs
are predominantly occurring nonlinear excitations in discrete nonlinear
systems. To understand this, we note that stable localized modes must
always be either below the band or in band gaps\cite{37}. So, the
discreteness in spatial variables can provide a favorable mechanism for
the formation and the stabilization of ILMs in discrete nonlinear systems
by introducing finite bandwidths, and consequently accessible band
edges. This, in turn increases the probability that the energy of a
localized self-organization in a discrete nonlinear system
will lie in the band gap. Again, the band width of a perfect linear
discrete system depends on the magnitude of the inter-site coupling
term. In a single  band model, if such  coupling is weak, we have a narrow
band. A discrete nonlinear system with narrow bands is called
anti-integrable. Such anti-integrable nonlinear systems are then expected
to sustain nonlinearity induced localized modes in the band gaps by the
above argument. There is indeed a mathematical proof of this in the
literature\cite{13,14}. Of course, it is not necessary to have
anti-integrable systems to have breathers. The existence of a breather
solution in the N-AL equation has been shown\cite{26}. In fact, in
contrast to continuous nonlinear systems, in any general discrete
nonlinear systems, particularly in nonintegrable systems, stationary
breathers are predominantly occurring ILMs.

 When localized states are formed below the lower band edge of a band,
they are unstaggered or symmetric localized states. These states are
symmetric under reflection through the center, and of course low energy
localized modes of the system\cite{26, 32}. Furthermore, these symmetric
localized states can have its peak at  a lattice site or in between two lattice
sites. The first one is called on-site peaked unstaggered localized
modes. The other one is called inter-site peaked unstaggered
localized modes. When localized states are formed above the upper
band edge of a band, they are staggered or antisymmetric localized
states\cite{32}. These states are antisymmetric under reflection and high
energy excitations of a system. In case of staggered localized states, we
analogously have  odd-parity Sievers-Takeno mode (ST) as well as
even-parity Page(P) mode\cite{45, 46, 47}. It is also
to be noted that staggered localized states have no analog in continuous
systems\cite{32}. Another kind of ILMs, called twisted localized modes
can be found in nonlinear lattices\cite{48, 49}. In this category also, we
can have unstaggered as well as staggered localized modes\cite{48,
49}. When these modes are stationary modes of the system, they are called
stationary localized states (SLS). We emphasize again that SLS of any
type, if they are true eigenmodes of nonlinear systems are also discrete
breathers\cite{11,12}. They may be called trivial breathers. 

Though it is possible to have in circumstances stationary ILMs in 
discrete integrable nonlinear systems, stationary ILMs are formed mostly
in nonintegrable nonlinear systems. We discuss here the stationary ILMs of 
SLS type. We know that stationary solitons of AL equations are  
examples of SLS in integrable nonlinear equations, and these are also
breather solutions of the same equation. We should, however not fail
to note that though these breather solutions are band edge states, their
widths are undetermined. On the other hand, the formation of SLS in
discrete nonlinear systems depends critically on two factors, the
inter-site hopping term which determines the width of bands in the
corresponding linear systems and the strength of the nonlinearity, which
determines the energy of the self organized localized formation. If the
first term is predominant, the nonlinearity can produce at best localized
modes near band edges. When localized states are formed near band edges,
they are weakly localized. In other words, they have large widths and
small amplitudes. Since, the movement of these localized modes does not
require large scale rearrangement in the lattice, such localized modes
can be made to move by applying small perturbing fields. As the movement
of any unstaggered localized state will not require an inversion of
orientation in any of sites, these states can easily move compared to its
staggered counterpart under small perturbation. Again, in case of
unstaggered localized states, inter-site peaked states will have larger
widths and smaller amplitudes compared to its on-site peaked
counterparts. So, inter-site peaked states can be made mobile easily by 
a small perturbation. In the other extreme where nonlinearity is strong,
strong localized modes having nonzero amplitudes only at a few sites are
formed. These are, of course high energy ILMs. Odd parity Sievers-Takeno
(ST) modes and even parity P modes in strongly anharmonic lattices are
examples of such strongly localized modes. Since these modes are formed
from the acoustic branch of anharmonic lattices, they appear above the
band and hence are staggered localized states. It is further found that ST
modes are unstable to an infinitesimal perturbation. However, this mode is
not destroyed by the perturbation. Instead, any perturbation makes it
move\cite{49}. On the other hand, P mode is stable and does not move by
small perturbations. The mobility difference of these modes can be
understood by the PN potential.  Because of the distribution of
amplitudes, ST modes are formed at the the maximum of the PN potential and
the P modes at the bottom of this potential\cite{50}. For the P mode to
move then we need enough energy to excite this mode above the PN
potential. Consequently, under a perturbation, not sufficient to take it
out of the well, this mode will remain immobile. On the other hand, ST
modes being at the maximum of the PN potential, no energy is needed to
take it out of the well. So, an infinitesimal perturbation can make it
mobile. The mobility difference of on-site and inter-site peaked
unstaggered localized modes to an infinitesimal perturbation can be
understood by the same argument. 

With this background, I plan to study here the formation of both
unstaggered and staggered stationary localized states in systems
described by IN-DNLS\cite{32}. To this end, I plan to examine the
dependence of the amplitude and width of the localized modes and also the
eigenfrequency of these modes on the nonintegrability parameter of the
equation. The energy of the localized modes are calculated from the
Hamiltonian. To the best of my knowledge, a rudimentary
asymptotic analysis of this problem is done using the lattice Green
function approach\cite{32,51}. For the detailed study, I plan to use the
discrete variational approach\cite{23, 42, 52}. In nonlinear dynamics, the
standard variational approach has been applied to continuous nonlinear
equations to study problems of nonlinear pulse propagation in optical
fibers, and to soliton dynamics in massive Thirring model, to mention a
few\cite{23, 53, 54, 55}. In the discrete variational approach, one
directly proceeds to search for discrete solutions of the coupled discrete
nonlinear evolution equations in a restricted subspace by imposing a
suitable ansatz for the solution\cite{23}. A procedure of averaging over
the discrete dimensions leads to either a set of coupled ODE's or a set of
coupled algebraic equations or both for the solution parameters. Therefore, 
this approach permits one to reduce the dimension of the problem from a
set of many coupled equations to generally a much smaller set of
equations, determined by the number of parameters in the ansatz to be
determined. Clearly, this method is advantageous when the number of
nonlinear equations is very large. This method has been applied to  DNLS,
for example  to study problems of beam steering in nonlinear waveguide
arrays\cite{23}, and also to understand  the formation and stability of
static and dynamical solitons in one dimensional systems and Cayley
trees\cite{42, 52}. We note in this context that equations like DNLS
describe the evolution of canonical coordinates of the canonical phase
space\cite{23, 56}. On the other hand, AL, N -AL and IN-DNLS in their
generic form describe the evolution of noncanonical coordinates in
noncanonical phase spaces\cite{24, 25, 26,32, 56}. Since, these
equations are derivable from Hamiltonians, the geometry of the dynamics is
automatically symplectic\cite{56}. The noncanonical symplectic structure
of the dynamics is manifested  in the structure of the Poisson
brackets\cite{8, 18, 26, 32, 36}. It is, however, to be noted that there
exists a global nonsingular coordinate transformation for these equations,
which transforms the noncanonical coordinates to canonical
coordinates\cite{36}. Therefore, these equations can also be
described by canonical coordinates with canonical Lagrangian and Poisson
brackets, having canonical symplectic structure\cite{36, 56}. I shall,
however proceed with the variational procedure with noncanonical
coordinates. I note that in Hamiltonian dynamics, the structure of the
Poisson bracket is incorporated in the Lagrangian\cite{36,57}. But, my
analysis is done with the appropriate functional, which is also obtainable
from the Lagrangian. So, the noncanonical symplectic structure of the
Poisson bracket does not pose any problem of finding SLS in IN-DNLS. The
other side of this analysis is the following. It shows how the effective
Lagrangian can be derived from the knowledge of the Hamiltonian and
constants of motion using the  analogous variational approach of finding
eigenvalues in standard Sturm-Liouville problems\cite{58}. In other words,
I shall also show that it is possible to set up the variational problem
for the determination of eigenvalues without the prior knowledge of the
Lagrangian.  Finally, I note that it has been seen in continuous nonlinear
equations that when the variational method is applied  to analyze solitary
wave dynamics, the solitary wave solutions may show instability in some
range of variational parameters. On the other hand, the correct dynamics
may not show at all such instability. So, the variational
method can produce false instabilities\cite{52, 53}. This consideration
also applies to discrete nonlinear evolution equations. However, I do not
encounter any undesired instability in my solutions, which can be ascribed
to the variational method. So, this aspect, even though important is not
dealt with here.

The organization of the paper is as follows. In the formulation section
below we present the basic equations to be studied. In the next section
we present a set of results, coming from one particular formulation. In
this section we also show that our formulation gives exact stationary
localized states of the AL equation. In the next section we present
another alternative formulation of the same problem. We then present the
corresponding results. Finally, we summarize our main results in the
summary section. Besides, this paper contains three important as well as
relevant Appendices.

\section{formalism}
\subsection{General derivation of  the nonlinear IN-DNLS equation and the
variational formulation of the corresponding eigenvalue problem }

We consider a dynamical system having 2N generalized noncanonical
coordinates, $\{\phi_{n},\ \phi^{\star}_{n}\},\ n\ =\ 1,\ \dots N $ in a
symplectic manifold\cite{56}. Let U and V be any two general dynamical
variables of the system. Any symplectic manifold has a natural Poisson
bracket structure, defined in terms of the inverse of the symplectic
structure function\cite{56}. So, we now define the following  noncanonical
Poisson bracket to characterize the manifold\cite{8, 32, 36}. 
\begin{widetext}
\begin{equation}
\{U,\ V \}_{\{\phi,\ \phi^{\star}\}}\ =\ i\  \sum_{n\ =\ 1}^{N}\
(\frac{\partial U} {\partial \phi_{n}}\  \frac{\partial V} 
{\partial \phi^{\star}_{n}}\ -\ \frac{\partial V} {\partial \phi_{n}}\
\frac{\partial U} {\partial \phi^{\star}_{n}})\ (1\ +\ \mu\
|\phi_{n}|^{2}).
\end{equation}

\noindent
We now consider the following Hamiltonian, ${\tilde{\rm{H}}}$.
\begin{equation}
{\tilde {\rm {H}}}  \ = \ -  \sum_{n}\
(\phi^{\star}_{n}\
\phi_{n+1}\ +\
\phi^{\star}_{n+1}\ \phi_{n})\ -\  2 \nu \ \sum_{n}
|\phi_{n}|^{2}\
+ \  2 \nu  \sum_{n} \ln{[1\ +\ |\phi_{n}|^{2}]};
\end{equation}
\end{widetext}
which is obtained from the original IN-DNLS Hamiltonian, $H$ through the 
transformations, $\phi_{n} \rightarrow  \sqrt{\mu}
\ \phi_{n},\ \ n\ \in Z\ {\rm {and}}\ \nu \rightarrow \frac{\nu}
{\mu}$\cite{32, 36}.
The corresponding Lagrangian ${\tilde{\mathcal{L}}}$ in the scaled
variables\cite{36, 57} is

\begin{equation}
{\tilde{\mathcal{L}}}\ =\ \frac{i} {2}\ \sum_{n}({\dot{\phi_{n}}}
\phi^{\star}_{n}\ - {\dot{\phi^{\star}_{n}}}
\phi_{n})\ \frac{\ln[1\ +\
|\phi_{n}|^{2}]} {|\phi_{n}|^{2}}\  -\ {\tilde {\rm {H}}} .
\end{equation}

The dynamical evolution of the n-th generalized coordinate, $\phi_{n}$,
can then be obtained by using Eq.(2.1) and Eq.(2.2).
\begin{eqnarray}
i\ \dot{\phi_{n}}\ &=&\ (1\ +\
|\phi_{n}|^{2})\ \frac{\partial {\tilde{\rm {H}}}} {\partial
{\phi^{\star}_{n}}}\nonumber\\
&=&\ - (1 + |\phi_{n}|^{2})(\phi_{n+1}\ +\ \phi_{n-1})\nonumber\\
&-&\ 2\ \nu |\phi_{n}|^{2}\ \phi_{n},
\end{eqnarray}
for $n\ \in \ Z$\cite{32, 36}. The other set of equations is obtained by
conjugation. The same equation can be obtained from the Lagrangian by
using the standard Lagrangian equations of motion. We note that under
the global gauge transformation, $\phi_{n} \rightarrow \phi_{n} e^{i
\alpha} $, Eqs.(2.2), (2.3) and (2.4) remain invariant. It can also
be shown from Eq.(2.4) that $ {\tilde{\mathcal{N}}}  =\  \sum_{n}\ln[1\ +\
|\phi_{n}|^{2}]$ is a constant of motion\cite{32}. We now assume that
$\phi_{n}\ =\ \lambda^{n}\ \Psi_{n}\ \exp{(- i\ \omega\ t)},\ n\ \in Z$
where $\lambda\ =\ \pm\ 1$. Furthermore, $\Psi_{n},\ n\ \in\ Z$ are
taken real\cite{32}. Then from Eq.(2.4), we get
\begin{eqnarray}
(\hat{\Omega}\ \hat{\Psi})_{n}\ &=&\ \omega\ \Psi_{n}\ +\ \lambda\ (1 +\
\Psi_{n}^{2})\ (\Psi_{n+1}\ +\ \Psi_{n-1})\nonumber\\ &+&\ 2\ \nu\
\Psi_{n}^{3}\ =\ 0.
\end{eqnarray}   
This is a nonlinear eigenvalue problem and its solutions give frequencies 
of stationary localized states of IN-DNLS equation\cite{32}. Introducing
the above ansatz for $\phi_{n},\ n\ \in Z$ in $\tilde{\mathcal{N}},\
{\rm{and}}\ \tilde{\rm{H}}$, we get
\begin{equation}
\tilde{\mathcal{N}}\ =\ \sum_{n}\ln[1 + \Psi_{n}^{2}],
\end{equation}
\begin{eqnarray}
\tilde{\rm{H}}\ &=&\ - 2\ \lambda\ \sum_{n} \Psi_{n}\ \Psi_{n+1} - 2 \nu
\sum_{n} \Psi_{n}^{2}\ +\ 2\ \nu\ \tilde{\mathcal{N}}\nonumber\\ 
&=&\ {\rm{\tilde {H}}_{0}}\ +\ 2\ \nu\ {\tilde{\mathcal{N}}}.
\end{eqnarray}
We define next
\begin{equation}
\tilde{\rm{F}}\ =\ \tilde{\rm{H}}\ -\ \Lambda\ \tilde{\mathcal{N}}
\end{equation}
\noindent
where $\Lambda$ is the Lagrange multiplier\cite{59}. Setting $\delta
{\tilde{\rm{F}}}\ =\ 0$, we get back Eq.(2.5),  when $\Lambda\ =\ \omega$. 
It is also important to note that the functional, $\tilde{\rm{ F}}$ can
also be obtained from ${\tilde{\mathcal{L}}}$ after introducing the
ansatz. In Appendix A, I plan to  discuss the importance of the
functional, $\tilde{\rm{ F}}$. 
 
\subsection{Variational approach with sech ansatz}
We first note that the system described by IN-DNLS equation, Eq.(2.4) has
lattice translational invariance. So, this system can
only form ILMs, arising from the competition between the localizing
nonlinearity and the dispersion from the inter-site hopping\cite{9}. As
the corresponding linear system is a discrete single band system, this
further enhances the propensity of formation of ILMs either below or above
the band. According to the theory of localization, any self-localized
state in 1-d systems will have exponential localization in the following
sense. The amplitude, $\Psi_{n}$ of the localized mode at the n-th site
will show exponential decay with $|n|$ for large values of
$|n|$\cite{37, 38}. We should also keep in mind that a modulus
function($|...|$) cannot appear in physical problem in its generic
form. This type of functions can only be obtained in any physical problem
in the asymptotic limit. Furthermore, when
$\nu\ = 0$ , Eq.(2.4) becomes the well known AL equation\cite{24, 25, 
32}. The one-soliton solution of Ablowitz-Ladik(AL) equation can either be
static or dynamic. For both cases, it has the sech profile, which
satisfies also the other requirement for localized states in
one dimension. So, we use the ansatz, $\Psi_{n}\ =\ \Phi \frac{1}
{\cosh\beta (n\ -\ x_{0})}\ ,\ n\ \in Z$. This ansatz has also been
used in the previous analysis\cite{32}. For on-site peaked and ST like
localized states, $x_{0}\ =\ 0$, and for inter-site peaked and P
like states, $x_{0}\ =\ \pm\frac{1} {2}$\cite{23, 45, 46, 47}. We further
write $\Phi^{2}\ =\ \Psi$. While $\beta^{-1}$
gives the half-width of localization, $\Phi$ denotes the maximum amplitude
of the states. Now, introduction of this ansatz in the functional
$\tilde{\rm{F}}$ makes it an algebraic function of the
parameters of the ansatz, 
\begin{equation}
\tilde{\mathcal{F}}(\Psi, \beta, \lambda, x_{0})\ =\ \tilde{\rm{H}}(\Psi,
\beta, \lambda, x_{0})\ -\ \Lambda\ \tilde{\mathcal{N}}(\Psi, \beta,
x_{0}), 
\end{equation} 
and we need to find relative extrema of $\tilde{\rm{F}}$ with respect to
variables, $\Psi$ and $\beta$\cite{59}. The finding of relative extrema
with respect to these two variables, $\Psi$ and $\beta$ means that $d
{\tilde{\rm{F}}}\ =\ 0$ should imply the following equations\cite{59}.
\begin{eqnarray}
\frac{\partial {\rm {\tilde{H}}_{0}}} {\partial \Psi\ \ }\  -\
\Lambda_{1}\ 
\frac{\partial \tilde {\mathcal{N}}} {\partial \Psi\ \ }\ =\ 0\\
\frac{\partial {\rm{\tilde{H}}_{0}}} {\partial \beta\ \ }\  -\
\Lambda_{1}\ 
\frac{\partial \tilde {\mathcal{N}}} {\partial \beta\ \ }\ =\ 0,
\end{eqnarray} 
where $\Lambda_{1}\ =\ \Lambda\ -\ 2 \nu$. For what follows next, we
assume that $\frac{\partial \tilde {\mathcal{N}}} {\partial \Psi}\ \not=\
0$. Then from Eqs.(2.10) and (2.11), we find that
\begin{equation}
\Lambda\ =\ \omega\ =\ 2 \nu\ + \frac{\frac{\partial {\rm 
{\tilde{H}}_{0}}}  {\partial \Psi\ \ }} {\frac{\partial \tilde
{\mathcal{N}}} {\partial \Psi\ \ }\ \ }
\end{equation}
and also
\begin{equation}
f(\Psi, \beta, \lambda, x_{0})\ =\ \{ {\rm{\tilde{H}}_{0}},\ 
{\tilde{\mathcal{N}}}\}_{\{\beta, \Psi \}}\ =\ 0.
\end{equation}
The other required equation is 
\begin{equation}
{\tilde{\mathcal{N}}}(\Psi, \beta, x_{0})\ =\ {\rm{C}}\ =\
{\rm{Constant}}.
\end{equation}
We note that we have three unknowns, namely $\Lambda, \Psi\ {\rm{and}}\
\beta$. But, we also have three independent equations to solve for these
unknowns. Hence, the problem is well-posed. 

\subsection{Calculation of ${\rm{\tilde{H}}_{0}}$ and 
${\tilde{\mathcal{N}}}$}
Introducing the expression of $\Psi_{n},\ n \in Z$ in ${\rm{\tilde{H}}_{0}}$ 
and ${\tilde{\mathcal{N}}}$ we get
\begin{equation} 
{\rm{\tilde{H}}_{0}}(\Psi, \beta, \lambda, x_{0})\ =\  - 2\ \lambda\
\Psi\ S_{1}(\beta, x_{0}) - 2 \nu\ \Psi\ S_{2}(\beta, x_{0}),
\end{equation}
\noindent
where
\begin{eqnarray}
S_{1}(\beta, x_{0}) &=& \sum_{n\ =\ - \infty}^{\infty} \frac{1}
{\cosh{\beta (n - x_{0})}\  \cosh{\beta (n + 1 - x_{0})}}\nonumber\\
S_{2}(\beta, x_{0}) &=& \sum_{n\ =\ -\infty}^{\infty} \frac{1}
{\cosh^{2}{\beta (n - x_{0})}}\nonumber
\end{eqnarray}
\begin{equation}
{\tilde{\mathcal{N}}}(\Psi, \beta, x_{0})\ = \ \sum_{n\ =\
-\infty}^{\infty} Y_{n}(\Psi, \beta, x_{0}) 
\end{equation}

\noindent
where
\[Y_{n}(\Psi, \beta, x_{0})\ =\ \ln{[1\ +\ \frac{\Psi} 
{\cosh^{2}{\beta (n - x_{0})}}]}.\]
\noindent 
To evaluate $S_{1}(\beta, x_{0}),\ S_{2}(\beta, x_{0})\ {\rm{and}}\
{\tilde{\mathcal{N}}}(\Psi, \beta, x_{0})$, we make  use of the
famous Poisson's sum formula\cite{23, 26, 28, 31, 43}  which reads
\begin{equation}
\sum^{\infty}_{n = -\infty} f(n\beta)\ =\ \frac{1} {\beta}
\int^{\infty}_{-\infty}
dy\ [1 + 2\
\sum^{\infty}_{s=1} \cos(\frac{2 \pi\ s\ y} {\beta})]\  f(y).   
\end{equation}
This application yields
\begin{eqnarray}
S_{1}(\beta, x_{0})\ \  =\ &\frac{2} {\sinh{\beta}},\ \ &\\
S_{2}(\beta, x_{0})\ =\ \frac{2} {\beta}\ +\ \frac{4} {\beta}\ \sum_{s\
=\ 1}^{\infty} &\Gamma_{s}(\beta, x_{0}),&\ \  
\end{eqnarray}

\noindent
where
\[\Gamma_{s}(\beta, x_{0})\  =\    \cos{2 \pi s x_{0}}\
\frac{\frac{\pi^{2} s} {\beta}} {\sinh{\frac {\pi^{2} s} {\beta}}}.\]
\begin{eqnarray}
\sqrt{\Psi (1 +\ \Psi)}\ {\frac{\partial \tilde{\mathcal{N}}} {\partial
\Psi\ \ }\ \ }\ =\ \frac{2} {\beta}\ &{\rm
{arc}}\sinh{\sqrt{\Psi}}&\nonumber\\ 
+\ \frac{2 \pi} {\beta}\ \sum_{s\ =\ 1}^{\infty} \cos{2 \pi s x_{0}}\
&T_{s}(\Psi, \beta),& 
\end{eqnarray}
\noindent
where
\[T_{s}(\Psi, \beta) = \frac{\sin{[ {\frac{2 \pi s} {\beta}} 
{\rm{arc}}\sinh{\sqrt{\Psi}}]}}
{\sinh{\frac{\pi^{2} s} {\beta}}}. \]

We now define a function, $f_{1}(\beta, \nu, \lambda, x_{0})$
\begin{equation}
f_{1}(\beta,\nu, \lambda, x_{0}) =\frac{\sinh{\beta}} {1 +\ \lambda\ 
\nu\ \frac{\sinh{\beta}} {\beta}\ S_{3}(\beta, \nu, \lambda, x_{0})},\ \ \
\end{equation}
\noindent
where
\[S_{3}(\beta, \nu, \lambda, x_{0}) = 1\ +\ 2\  \sum_{s\ =\
1}^{\infty} \cos{2 \pi s x_{0}}\  \frac{\frac{\pi^{2} s} {\beta}}
{\sinh{\frac{\pi^{2} s} {\beta}}}.\]
Now, with this definition, we have
\begin{eqnarray}
{\rm{\tilde{H}}_{0}}\ &=&\ - 4\ \lambda\ \frac{\Psi} {f_{1}(\beta,
\nu, \lambda, x_{0})}\\
\frac{\partial {\rm {\tilde{H}}_{0}}} {\partial \Psi\ \ }\ &=&\ -\ 
\frac{4\ 
\lambda} {f_{1}(\beta, \nu, \lambda, x_{0})}\\
\frac{\partial {\rm{\tilde{H}}_{0}}} {\partial \beta\ \ }\ 
&=&\ -\ 4\ \lambda\ \Psi\  \frac{\partial {\frac{1} {f_{1}(\beta, \nu,
\lambda, x_{0})}}} {\partial \beta\ \ \ \ \ \ \ \ \ \ \ }.
\end{eqnarray}
\begin{widetext} 
\noindent
Again from Eq.(2.20), we have
\begin{equation}
{\tilde{\mathcal{N}}}(\Psi, \beta, x_{0})\ =\ \frac{2} {\beta}\ 
({\rm{arc}}\sinh{\sqrt{\Psi}})^{2}\ 
+\ 4 \sum^{\infty}_{s\ =\ 1} \cos{2
\pi s x_{0}}\ \frac{\sin^{2}{({\frac{\pi\ s}
{\beta}}\ {\rm{arc}}\sinh{\sqrt{\Psi}})}} {s\ \sinh{\frac{\pi^{2} s}
{\beta}}},
\end{equation}
\noindent
and from Eq.(2.25) we in turn get
\begin{eqnarray}
\frac{\partial \tilde {\mathcal{N}}} {\partial \beta\ \ }\ &=&\ -\
\frac{2}
{\beta^{2}}\ ({\rm{arc}}\sinh{\sqrt{\Psi}})^{2}\nonumber\\ 
&-&\ \frac{4\ \pi\  {\rm{arc}}\sinh{\sqrt{\Psi}}} {\beta^{2}}\ 
\sum^{\infty}_{s\ =\ 1} \cos{2\pi s x_{0}}\ 
\frac{\sin{({\frac{2\ \pi\ s}
{\beta}}\ {\rm{arc}}\sinh{\sqrt{\Psi}})}} {\sinh{\frac{\pi^{2} s} 
{\beta}}}\nonumber\\
&+&\ \frac{4\ \pi^{2}} {\beta^{2}}\ \sum^{\infty}_{s\ =\ 1} \cos{2
\pi s x_{0}}\ \frac{\sin^{2}{({\frac{\pi\ s}
{\beta}}\ {\rm{arc}}\sinh{\sqrt{\Psi}})}} {\sinh{\frac{\pi^{2} s}
{\beta}}}\ \coth{\frac{\pi^{2} s} {\beta}}.
\end{eqnarray}
\noindent
The calculation of Eq.(2.25) is given in Appendix B.
\end{widetext}
In our variational formulation, in principle $x_{0}$ is another parameter
to be determined from the extrema of the functional, $\tilde{\rm{F}}$
(Eq.(2.8)). Now, the extremization of  $\tilde{\rm{F}}$ with $x_{0}$
inclusive will yield along with Eqs.(2.10) and(2.11), the following
equation. 
\begin{equation}
\frac{\partial {\rm {\tilde{H}}_{0}}} {\partial x_{0}\ \ }\  -\
\Lambda_{1}\
\frac{\partial \tilde {\mathcal{N}}} {\partial x_{0}\ \ }\ =\ 0.
\end{equation}

\noindent
But, from Eqs.(2.21), (2.22) and (2.25), it can be easily proved that as
$0\ \le\ |x_{0}|\ < 1$,  $x_{0}\ \ =\ 0,\ \  \pm \frac{1} {2}$.

\section{The variational formulation with ${\tilde{\mathcal{N}}}$ constant
: results and discussion}
\subsection{Ablowitz-Ladik limit}
In the Ablowitz-Ladik limit, $\nu\ =\ 0$. To probe this limit, we evaluate
relevant functions and their derivatives along the curve $\Psi\ =\
\sinh^{2}{\beta}$. Along this curve, from Eqs.(2.20), (2.25) and (2.26) we
have 
\begin{eqnarray}
{\tilde{\mathcal{N}}}(\Psi, \beta, x_{0})\ &=&\ 2\ \beta\\
\frac{\partial \tilde {\mathcal{N}}} {\partial \beta\ \ }\ &=&\ -\ 2\\
\frac{\partial \tilde{\mathcal{N}}} {\partial \Psi\ \ }\  &=&\
\frac{2} {\sinh{\beta}\ \cosh{\beta}}.
\end{eqnarray}
Since, $\nu\ =\ 0$ in this case, we also have  from Eqs.(2.21) to (2.24)
\begin{eqnarray}
\frac{\partial {\rm {\tilde{H}}_{0}}} {\partial \Psi\ \ }\ &=&\ -\
\frac{4\ \lambda} {\sinh{\beta}}\\
\frac{\partial {\rm{\tilde{H}}_{0}}} {\partial \beta\ \ }\
&=&\ \ 4\ \lambda\ \cosh{\beta}. 
\end{eqnarray}
We find from Eqs.(3.2) to (3.5) that $ f(\Psi, \beta, \lambda, x_{0})\
=\ \{ {\rm{\tilde{H}}_{0}},\ {\tilde{\mathcal{N}}}\}_{\{\beta, \Psi \}}\
=\ 0.$ Furthermore, from Eqs.(2.12), (3.3) and (3.4) we get $\omega\
=\ -\ 2\ \lambda\ \cosh{\beta}$. The energy, ${\tilde{\rm{E}}}\ =\
{\tilde{\rm{H}}}\ =\ -\ 4\ \lambda\ \sinh{\beta}$. Due to positive
semi-definiteness of $\tilde{\mathcal{N}}$, we get from Eq.(3.1) that
$\beta$ should also be positive semidefinite. This
is consistent with the one-soliton solution of Ablowitz-Ladik equation.  

We now consider the case of $\nu\ \not=\ 0$. For convenience, we define
\begin{equation}
g(\beta, x_{0})\ =\ \frac{1} {\beta}\
[ 1\ +\ 2\  \sum_{s\ =\
1}^{\infty} \cos{2 \pi s x_{0}}\  \frac{\frac{\pi^{2} s} {\beta}}
{\sinh{\frac{\pi^{2} s} {\beta}}}]
\end{equation}
\noindent
Along the line $\Psi\ =\ \sinh^{2}{\beta}$, we find that
\begin{eqnarray}
f(\Psi, \beta, \lambda, x_{0})\ & =&\ \{{\rm{\tilde{H}}_{0}},\
{\tilde{\mathcal{N}}}\}_{\{\beta, \Psi \}}\nonumber\\ 
=\ -\ 8\ \nu\
g(\beta, x_{0})\
&\tanh{\beta}&\ \frac{d \ln{A_{0}(\beta, x_{0})}} {d \beta\ \ \ \ \ \ \ \
\ \  \ \ \ \ \ }
\end{eqnarray}
where $A_{0}(\beta, x_{0}) = \sinh{\beta}\ g(\beta, x_{0})$. 
When $\beta  \rightarrow  0$, $\tanh{\beta}\  \frac{d \ln{A_{0}(\beta, 
x_{0})}} {d \beta } \rightarrow \frac{\beta^{2}} {3}$, and consequently 
$f(\Psi, \beta, \lambda, x_{0}) \sim\ -  \frac{8 \nu} {3} \beta^{2}$,
provided $\nu$ is finite. So, when
$(\nu\ \beta^{2})\ \sim\ o (1),\ \Psi\ =\ \sinh^{2}{\beta}$ is an
asymptotic solution of a localized state with a large width and a small
amplitude. Eigenvalue, $\omega$ and energy, $\tilde{\rm {E}}\
=\tilde{\rm{H}}$ of these localized states are
\begin{eqnarray}
\omega\ &=&\ 2\ \nu\ -\ 2(\lambda\ +\ \nu\ A_{0})\ \cosh{\beta}\nonumber\\ 
&\sim&\ -\ 2\ \lambda\ -\ (\lambda\ +\ \frac{4 \nu} {3})\ \beta^{2},\\
{\rm{and}}\nonumber\\
\tilde{\rm{E}}\ &=&\ -\ 4\ \lambda\ \beta\ -\ 4\ \nu\
\beta\ [\frac{A_{0}\ \sinh{\beta}} {\beta}\ -\ 1]\nonumber\\ 
&\sim&\ -\ 4\ \lambda\ \beta\ -\ \frac{4\ \nu} {3}\ \beta^{3}. 
\end{eqnarray} 
\noindent
So, according to this asymptotic analysis, when $\nu\ \not =\ 0$, the
nonintegrability parameter, $\nu$ and the width parameter $\beta$ of the
SLS are not independent of each other.

\subsection{Stationary localized states from IN-DNLS}
We now consider various mathematical aspects of the formation of
stationary localized states in IN-DNLS. We consider first Eq.(2.14) along
with Eq.(2.25). We restrict ourselves to $\beta  \ge 0$, which is
necessary to keep $\Psi$ positive semi-definite. Furthermore, in the
following analysis, we assume that $\beta \le  1$. In this situation, we
can ignore infinite sums in Eq.(2.21) and in Eq.(2.25). Due to this
approximation, Eq.(2.25) yields
$\Psi = \sinh^{2}{\alpha  \sqrt{\beta}}$ where $\alpha$ is a constant,
as required by Eq.(2.14). Since, the right hand side of Eq.(2.14) is
taken to be a number constant, ${\rm{C}} =  2.0\  \alpha^{2} $, we
have $\frac{d \tilde {\mathcal{N}}} {d \beta\ \ }  =  0$ irrespective of
the value of $\beta$. This, in turn gives
\begin{equation}
\frac{d \Psi} {d \beta}\ =\ -\ \frac{\frac{\partial \tilde {\mathcal{N}}}
{\partial
\beta\ \ }} {\frac{\partial \tilde{\mathcal{N}}} {\partial \Psi\ \ }\ \ }.
\end{equation} 
Now introducing Eq.(3.10) in Eq.(2.13), we get $\frac{d
{\rm{\tilde{H}}_{0}}}
{d \beta\ \ }  = 0$. In other words, permissible values of $\beta$
are determined from the extrema of ${\rm{\tilde{H}}_{0}}$ as a
function of $\beta$. From the functional dependence of 
${\rm{\tilde{H}}_{0}}$, and $\Psi$ on $\beta$, we ultimately get 
\begin{eqnarray}
g_{1}(\alpha, \beta)\ &=&\ \frac{\beta} {\sinh{\beta}}\  \cosh{\beta}\ 
\frac{\tanh{\alpha \sqrt{\beta}}} {\alpha \sqrt{\beta}}\ - 1\nonumber\\
g_{2}(\alpha, \beta)\ &=&\ 1 - \frac{\tanh{\alpha \sqrt{\beta}}} {\alpha
\sqrt{\beta}}\nonumber\\
\nu \lambda\ &=&\ \frac{\beta} {\sinh{\beta}}\ \ \frac{g_{1}(\alpha,
\beta)} {g_{2}(\alpha, \beta)}.
\end{eqnarray}
We note that for a given value of the parameter, $\alpha$, $\beta$ is
determined by the nonintegrability parameter, $\nu$.
Furthermore, Eq.(3.11) yields two positive values of
$\beta$ as roots, if two conditions, namely  $\nu  \lambda \ge  0$ and
and $|\nu| <   |\nu_{\rm{critical}}|$. The
behavior of the smaller root ($\beta_{s}$) as a function of $\nu$ for
$\lambda\ =1 $ and $\alpha\ =0.5\ {\rm{and}}\ 0.25$ are shown in Fig.1 
\begin{figure}
\parindent 0.3in
\includegraphics{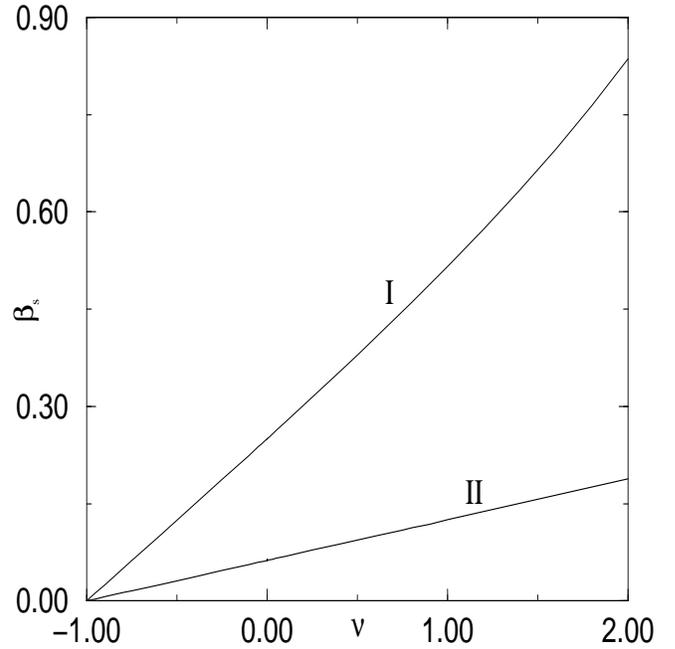}
\caption{This figure shows the variation of the smaller root, $\beta_{s}$
of Eq.(3.11) as a function of the nonintegrability parameter, $\nu$. Since
$\lambda = 1$, these states are unstaggered stationary localized
states. Curve I : $\alpha$ = 0.5 and Curve II : $\alpha$ = 0.25.}
\end{figure}
\begin{figure}
\parindent 0.3in
\includegraphics{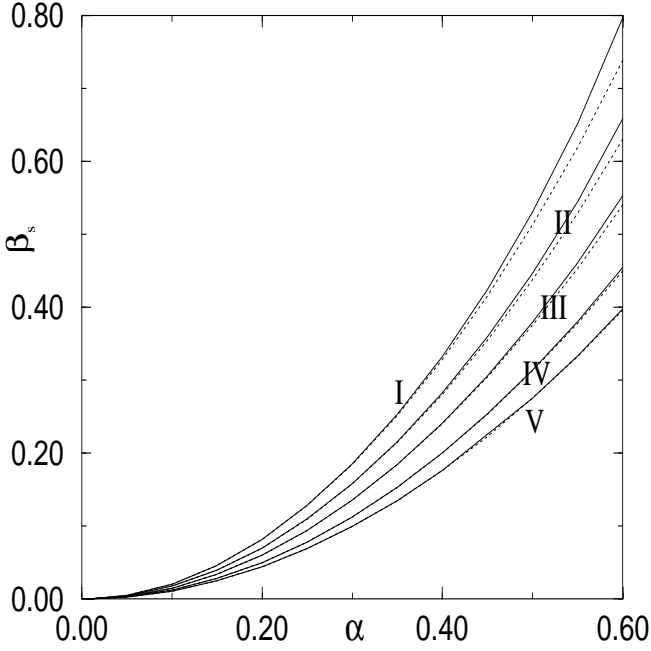}
\caption{This figure shows the variation of the smaller root,
$\beta_{s}$ of Eq.(3.11) as a function of the parameter, $\alpha$ for
various values of the nonintegrability parameter, $\nu$. Since
$\lambda = 1$, these states are unstaggered stationary localized
states. Curve I : $\nu$ = 1.05,   Curve II : $\nu$ = 0.75,
Curve III : $\nu$ = 0.5,  Curve IV : $\nu$ = 0.25, and
Curve V : $\nu$ = 0.10. Each curve is associated with a dotted curve
which shows the variation of $\alpha^{2} (1 + \nu \lambda)$ as a function
of $\alpha$ for the corresponding value of $\nu$.}
\end{figure}
In Fig.2 we present the variation of $\beta_{s}$ as a function of the
parameter, $\alpha$ for various values of $\nu \ge 0$. It should be noted
from these figures that $\beta_{s} \le  1 $ for these values of $\alpha$
and the chosen interval of $\nu$. So, the neglect of infinite sums in
Eqs.(2.21) and (2.25) is justified. It is a simple exercise to see from
Eq.(3.11) that when $|\nu| \rightarrow 0,\ \beta_{s}  \rightarrow
\alpha^{2}$. Then, for small values of $\nu$
the asymptotic solution is the AL stationary localized state solution. This 
is a very important result. This asymptotic analysis reveals that this
stationary localized state solution of IN-DNLS continuously moves to the
AL stationary localized state solution when $\nu  \rightarrow  0$ from
either side. It is further important to note from Fig.2 that for $\alpha
\ll  1$,\  we have  $\alpha\ \approx\ \sqrt{\frac{\beta_{s}} {1.0\ +\ \nu\
\lambda}},\  \nu\ \lambda\ \ge\ 0$. Consequently, $\sqrt{\Psi}\ \approx\ 
\sinh{\frac{\beta_{s}} {\sqrt{1.0\ +\ \nu\ \lambda}}}$.  But, as for this
range of argument, $\sinh{x}\ \approx\ x$,  we have $\sqrt{\Psi}\ \sim\
\frac{\beta_{s}} {\sqrt{1.0 \ +\ \nu\ \lambda}}$. This agrees with the
existing asymptotic analysis\cite{32}.

The variation of the large root, $\beta_{l}$ as a function of $\nu
\lambda$ for various values of $\alpha$ is shown in Fig.3. 
\begin{figure}
\parindent 0.25in
\includegraphics{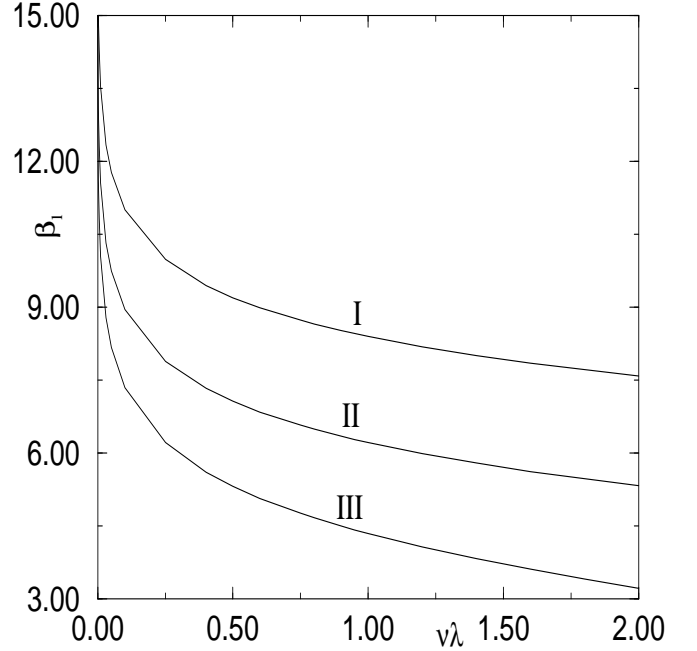}
\caption{This figure shows the variation of the larger root, $\beta_{l}$
of Eq.(3.11) as a function of $\nu \lambda$. If $\lambda = 1$, these
states are then unstaggered stationary localized states. Curve I
: $\alpha$ = 0.1 and Curve II : $\alpha$ = 0.25, and Curve III : $\alpha =
0.50$.}
\end{figure}
\begin{figure}
\parindent 0.3in
\includegraphics{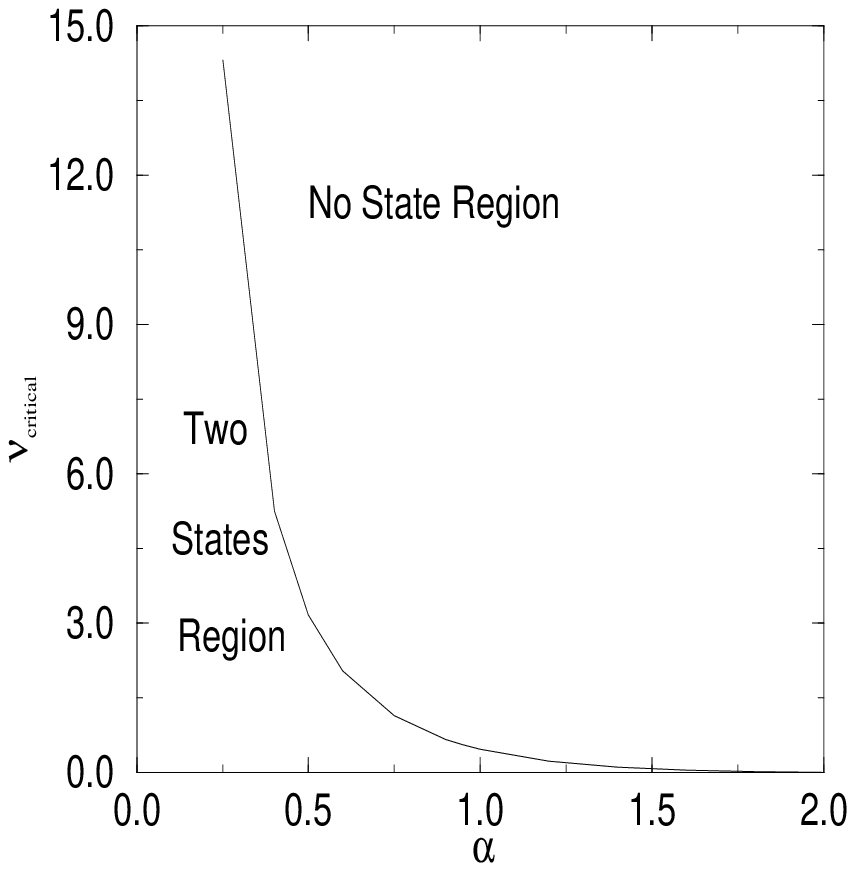}
\caption{This figure shows the variation of $\nu_{\rm{critical}}$ as a
function of the parameter, $\alpha$ for unstaggered stationary localized
states, $\lambda =1 $. $\nu_{\rm{critical}}$ is obtained from
Eqs.(3.11) and (3.12) in the text. Note that the curve separates the two
states region from the no state region.}
\end{figure}
Again, by comparing Fig.1 and Fig.3, we see that as $\nu$ increases, the
large root, $\beta_{l}$, of Eq.(3.11) decreases from $\infty$, while the
other root, $\beta_{s}$  increases from zero. So, for a given $\alpha$,
the value of $\nu_{critical}$ is determined by the inflection point of
${\rm{\tilde{H}_{0}}}$. This then implies that equations to solve for
$\beta_{critical}$, and  $\nu_{critical}$ are obtained by setting both
$\frac{d {\rm{\tilde{H}_{0}}}} {d \beta\ \ }  =  0$ and $\frac{d^{2} 
{\rm{\tilde{H}_{0}}}} {d \beta^{2}\ \ }  =  0$. While the first condition
gives Eq.(3.11), the second condition yields Eq.(3.12) as shown below.
\[g_{3}(\beta)\ = \ \frac{\tanh{\sqrt{\beta}}} {\sqrt{\beta}}\ \
{\rm{and}}\ \ 
g_{4}(\beta) \ =\ \frac{\beta} {\sinh{\beta}},\]
\begin{eqnarray}
g_{5}(\alpha, \beta) \ & =&\  1\ -\ g_{3}(4 \alpha^{2} \beta) - 4\
g_{3}(4 \alpha^{2} \beta) (1 - \ g_{3}(\alpha^{2} \beta)),\nonumber\\
g_{6}(\alpha, \beta)\ &=&\  1 - g_{3}(4 \alpha^{2}\ \beta)\ (1 + 4
\ \cosh{\beta}\ g_{4}(\beta)),\nonumber\\ 
g_{7}(\alpha, \beta)\ &=&\  g_{4}(\beta)\ g^{2}_{3}(\alpha^{2} \beta)\
(\beta^{2}\ +\ 2\ g^{2}_{4}(\beta)),\nonumber\\
g_{8}(\alpha, \beta)\ &=&\ -2 \nu\  \frac{\alpha^{2}} {\beta^{2}}\
\cosh{2 \alpha \sqrt{\beta}}\ g_{5}(\alpha, \beta),\nonumber\\
g_{9}(\alpha, \beta)\ &=&\ - 2\ \lambda\ \frac{\alpha^{2}} {\beta^{2}}\
\cosh{2 \alpha\ \sqrt{\beta}}\ g_{4}(\beta)\ g_{6}(\alpha,
\beta),\nonumber\\
g_{10}(\alpha, \beta)\ & =&\ - 4\ \lambda\
\frac{\alpha^{2}} {\beta^{2}}\ \cosh^{2}{\alpha \sqrt{\beta}}\
g_{7}(\alpha, \beta),\nonumber \\
g_{8}(\alpha, \beta)\ &+&\ g_{9}(\alpha, \beta)\ +\ g_{10}(\alpha, \beta)\
=\ 0.
\end{eqnarray}
We again note that $\lambda  =  \pm 1$ and $\nu$ in Eq.(3.12) is given by
Eq.(3.11). Of course, $\nu \lambda$ is positive. We find from
Eq.(3.12) that when $\alpha  \rightarrow  0$, $\nu_{\rm{critical}}
\rightarrow  \infty$. Again, when $\alpha \gg  1$, $\nu_{\rm{critical}}
\sim 0$. The functional dependence of $\nu_{crititcal}$ on $\alpha$ is
shown in Fig.4.

The other important case is where  $\nu  \lambda  < 0$. This means
that either we have an unstaggered state with $- \nu$ or a staggered state
with $+ \nu$. In this case if $|\nu \lambda|  >  1 $, both roots of
Eq.(3.11) are negative. Inasmuch as  ${\tilde{\mathcal{N}}}(\psi, \beta,
x_{0})$ is positive semi-definite, it is easy to see from Eq.(2.25) that
this is not permissible. For this case, from Eq.(3.11) as expectedly we
obtain that when $\nu \lambda \rightarrow  -1 +,\ \beta  \rightarrow 
0$, and when $\nu \lambda  \rightarrow  0 -,\ \beta  \rightarrow 
\alpha^{2}$. See both Fig.1 and Fig.5. 
\begin{figure}
\parindent 0.37in
\includegraphics{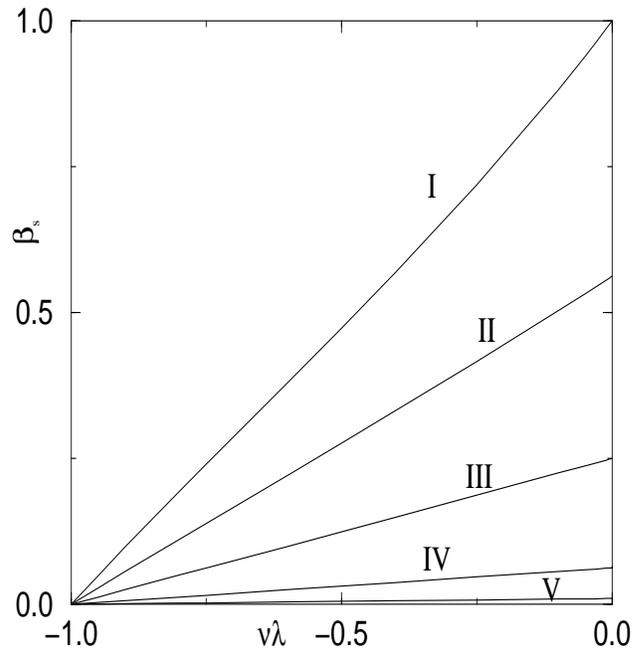}
\caption{This figure shows the variation of the smaller root, $\beta_{s}$
of Eq.(3.11) as a function of $\nu \lambda$ for $\nu \lambda < 0$
for various values of the nonintegrability parameter, $\alpha$. For
$\lambda = - 1$, these states are staggered stationary localized
states. Curve I : $\alpha$ = 1.0,   Curve II : $\alpha$ = 0.75,
Curve III : $\alpha$ = 0.5,  Curve IV : $\alpha$ = 0.25,
and  Curve V : $\alpha$ = 0.10.}
\end{figure}
\begin{figure}
\parindent 0.3in
\includegraphics{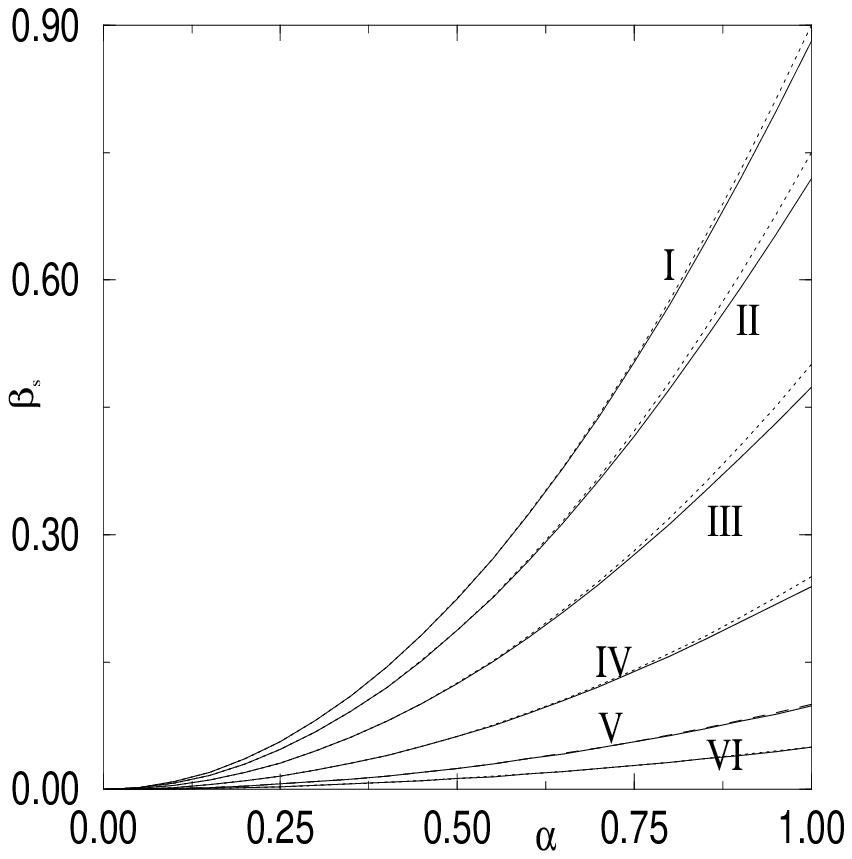}
\caption{This figure shows the variation of the smaller root,
$\beta_{s}$ of Eq.(3.11) as a function of the parameter, $\alpha$ for
various values of the nonintegrability parameter, $\nu$. Since
$\lambda = - 1$, these states are staggered stationary localized
states. Curve I : $\nu$ = 0.1,   Curve II : $\nu$ = 0.25, Curve III
: $\nu$ = 0.5,  Curve IV : $\nu$ = 0.75, Curve V : $\nu$ = 0.90 and Curve
VI : $\nu$ = 0.95. Each curve is also associated as in Fig.2 with a dotted
curve which shows the variation of $\alpha^{2} (1 + \nu \lambda)$ as a
function of $\alpha$ for the corresponding value of $\nu$.}
\end{figure}
Inasmuch as for $\alpha  \le  1 $, permissible values of $\beta_{s}  \le
1$, the neglect of infinite sums in Eqs.(2.21) and (2.25) is again well
justified. The variation of $\beta_{s}$ as a function of $\nu$ for
$\alpha\ = 1.0,\ 0.75,\ 0.5,\  0.25,\  {\rm{and}}\  0.10$ are shown in
Fig.5. It is seen from Fig.6 that when $\nu  \lambda  <  0,  \  \alpha \
=\ \sqrt{\frac{\beta_{s}} {1.0 \ +\ \nu\ \lambda}}$ is a very good 
approximation\cite{32}. Another important aspect is in Fig.7, which  shows
that for a given $\nu >  0$, the staggered SLS $(\lambda  =  - 1)$ has
larger width than the corresponding unstaggered SLS $(\lambda  =1)$. So,
the SLS for $\nu \lambda  <  0$ are  basically localized states with large
widths and small amplitudes.
\begin{figure}
\parindent 0.3in
\includegraphics{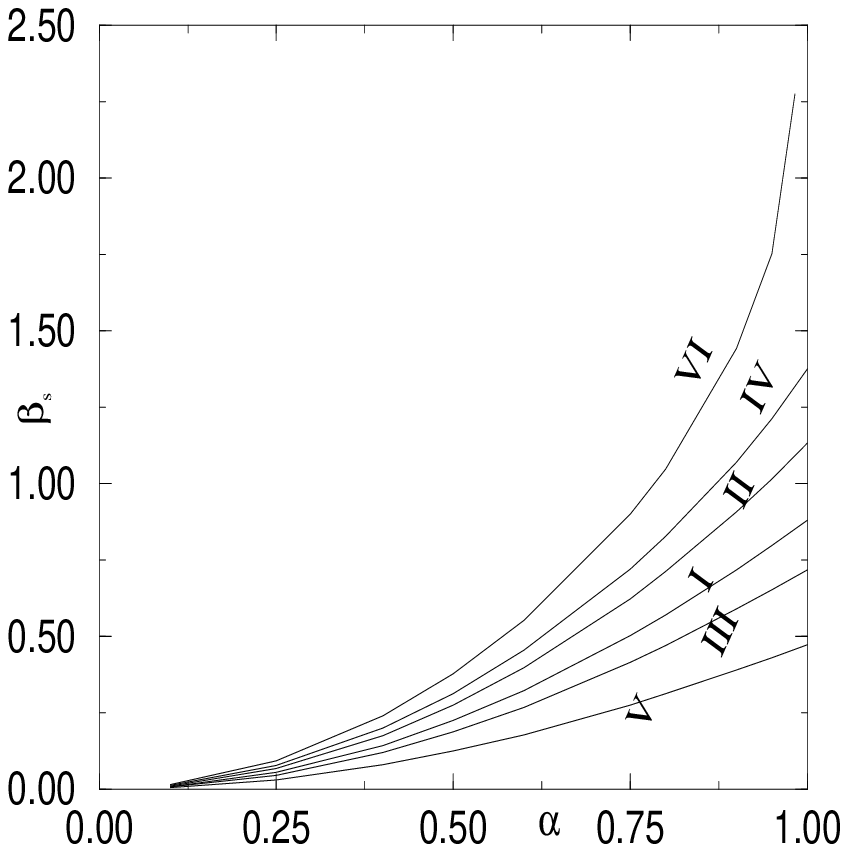}
\caption{ This figure compares the variation of the smaller root,
$\beta_{s}$ of Eq.(3.11) for both unstaggered ($\lambda$ = 1) and   
staggered ($\lambda$ = -1) stationary localized states as a function of
the parameter, $\alpha$ for various  values of the nonintegrability
parameter, $\nu$. Curve I : $\nu \lambda$ = - 0.1,  Curve II : $\nu   
\lambda$ = 0.1, Curve III : $\nu \lambda$ = - 0.25,
Curve IV : $\nu \lambda$ = 0.25,  Curve V : $\nu \lambda$ = - 0.5,
and Curve VI : $\nu \lambda$ = 0.5.}
\end{figure}
As for eigenvalues of these stationary localized states,  
introducing Eqs.(2.20) and (2.23) into Eq.(2.12) and using the
same approximation as used for finding the roots of Eq.(3.11), we obtain
that
\begin{eqnarray}
\omega\ & =&\ -\  2 \nu\ \  [ \frac{\sinh{2 \alpha \sqrt{\beta}}} {2
\alpha \sqrt{\beta}}\ -\ 1]\nonumber\\
& - &\ 2\ \lambda\  \frac{\beta} {\sinh{\beta}}\
\ \frac{\sinh{2 \alpha \sqrt{\beta}}} {2 \alpha \sqrt{\beta}}.
\end{eqnarray} 
The energy of these stationary states is given by
\begin{eqnarray}
{\tilde{\rm{E}}}\ & =&\ {\tilde{\rm{H}}}\ =\ {\rm{\tilde{H}}_{0}}\ +\ 2 \
\nu\ {\tilde{\mathcal{N}}}\nonumber\\ 
&=& \ - 4\ \lambda\ \frac{\sinh^{2}{\alpha\
\sqrt{\beta}}} {\sinh{\beta}}\nonumber\\ 
&-&\ 4\ \nu\ \alpha^{2}\ [ \frac{\sinh^{2}{\alpha \sqrt{\beta}}}
{\alpha^{2} \beta}\ -\ 1]. 
\end{eqnarray}
$\beta$ in Eqs.(3.13) and (3.14) is the root of Eq.(3.11). For $\beta  = 
\beta_{s}$ and $\alpha$ not too large, these equations are well
justified. We already noted that when $\nu  =  0,\  \beta_{s} =
\alpha^{2}$. Furthermore,  $|\nu \lambda|  \ll  1$ and also $\alpha \ll 
1,\  \beta_{s} \rightarrow  \alpha^{2}$. We obtain the respective
limiting results for these cases  from Eqs.(3.13) and (3.14).

\subsection{Stability and position  of stationary localized states of
IN-DNLS}
 We now discuss the issue of stability of these stationary
localized states. We note first that when $\nu  =  0$, the resulting
nonlinear equation is the AL equation, which has both
unstaggered and staggered stationary localized states. These are
basically band edge states. Our variational calculation correctly produces
these states of the AL equation, by letting $\beta_{s} \rightarrow
\alpha^{2}$ as $|\nu| \rightarrow 0$. See Fig.1 and Fig.5.  Over and above
it suggests another state for which  $\beta_{l} = \infty$. This can be
easily seen in  Eq.(3.11). It is again seen from 
Fig.1 and Fig.2 that for any $\alpha$, introduction of any
$\nu$, however small, with  $\nu \lambda  >  0$  makes
$\beta_{s} >  \alpha^{2}$. We observe that $\beta^{-1}_{s}$ gives the
half-width of the localized state. So, for these localized states, the 
half-width reduces with increasing $\nu$. Since for $\lambda  = 1$ and
$\nu  > 0$ $(\nu  \lambda  > 0)$ implies that the on-site nonlinear
trapping potential is attractive, any positive enhancement of $\nu$ should
reduce the half-width of the SLS by effectively reducing the inter-site
hopping. Whether a given $\nu$ defines an attractive or a repulsive
potential also depends upon the value of $\lambda$. So, the above
argument will hold good whenever $\nu \lambda > 0$. When $\nu 
\lambda$ is positive, unstaggered stationary localized states
characterized by $\beta_{s}$ are stable. On the other hand, for $\nu = 0$,
if there is any stationary localized state corresponding to
$\beta_{l}  =  \infty$, it is a state with a peak of infinite height at a
given site with a half-width of a few sites.  Again, we see from Fig.3
that for any $\alpha$, when $\nu \lambda$ increases, $\beta_{l}$
decreases. This means that the half-width increases. On the other hand,
the introduction of $\nu$ with $\nu \lambda >  0$ should reduce the
half-width, as our argument suggests.  Hence, stationary localized
states corresponding to $\beta_{l}$ are unstable. These states, if exist
in this system, will be unstable towards perturbation. 

Consider next the case where $\nu \lambda  < 0$. In this
case, we have either staggered localized states for positive $\nu$, or
unstaggered localized states for negative $\nu$. First of all,  there is
only one set of stationary localized states. Furthermore,
$0 \le \beta_{s}  \le  \alpha^{2}$ for  $ -1 \le\ \nu\ \lambda  \le
0$. See Fig.5. Since, for $\nu > 0$, staggered localized states
are stabilized by increasing the half-width (see Fig.7), states
characterized by $\beta_{s}$ are stable. For $\nu$ negative and $\lambda 
= 1$, or $\nu > 0$  and $\lambda = - 1$, the on-site nonlinear potential
is repulsive. So, the expansion of the half-width with decreasing $\nu$
is energetically favorable (see Fig.7).

So far our analysis did not include the effect of $x_{0}$, the position of
the peak on the formation of stationary localized states and their
stability. But, this is also an important part of the problem. However,
even a semi-rigorous investigation of  this problem in this formulation
requires the analytical solution of  $\Psi$ as exactly as possible from
Eq.(2.25). As there is no simple analytical way of
solving Eq.(2.25) for $\Psi$, one can take recourse to approximation
methods like, the method of  {\it{successive substitutions}}\cite{60}. I
shall describe in Appendix C how this method can be used to get
approximate dependence on $x_{0}$, of $\beta$, $\omega$ and $E$ of the
SLS. Of course, the other possibility is to find real positive
roots of Eq.(2.25) graphically. Inasmuch as the exact analytical solution
of $\Psi$ as a function of $\beta$ is difficult in this approach, we shall
not follow the present line of investigation further. On the contrary, we
shall show next how the exact dependence of the parameter, $\beta$, the
frequency $\omega$ and the energy $E$ of the SLS on $x_{0}$ can be
obtained by a rational alternation in the variational procedure.

\section{The variational formulation with  ${\rm{\tilde{H}}_{0}}$ constant
: results and discussion}
 
Since, ${\rm{\tilde{H}}}$ and $\tilde{\mathcal{N}}$ (Eq.2.2) and
Eq.(2.6) respectively) are two constants of motion, from the expression
of ${\rm{\tilde{H}}}$, we see that  ${{\tilde{H}}_{0}}$ (Eq.(2.7)) is also
a constant of motion. So, we reformulate in this section our variational
problem in which, ${{\tilde{H}}_{0}}$, in lieu of  $\tilde{\mathcal{N}}$
is taken to be the number constant. In this modified variational approach,
we take ${\rm{\tilde{F}}}\ =\ \Lambda_{2}\ {\rm{\tilde{H}_{0}}}\ +\
2\ \nu\ \tilde{\mathcal{N}}$, where $(\Lambda_{2}\ -\ 1)$ is the Lagrange
multiplier. This modified approach also yields Eq.(2.12) for the
eigenvalue, $\omega$ in Eq.(2.5). One other equation required to solve for
one of the two unknowns, namely $\beta$ and $\Psi$ is given by
Eq.(2.13). These results are also derived in  Appendix A. From
Eq.(2.22) we find that
\begin{equation}
\Psi(\beta, \nu, \lambda, x_{0})\ =\ a\ f_{1}(\beta, \nu, \lambda, x_{0})
\end{equation}  
where $a$ is a number constant, yields  ${\rm{\tilde{H}}_{0}}  
=  -  4  \lambda  a$, which is again a number constant. Most importantly,
in this formulation $\Psi$ is determined explicitly in terms of $\beta$
within a multiplicative number constant, $a$.  Inasmuch as $\Psi$ is
positive semi-definite by definition, the sign of this constant should be
such that $a  f_{1}$ is positive semi-definite. Furthermore, in this
approach  Eq.(2.22) yields $\frac{d {\rm{\tilde{H}}_{0}}} {d \beta\ \ }\
=\ 0$, irrespective of $\beta$. This in turn gives  
\begin{equation}
\frac{d \Psi} {d \beta}\ =\  -\ \frac{\frac{\partial {\rm{\tilde{H}}_{0}}}
{\partial \beta\ \ } } {\frac{\partial {\rm {\tilde{H}}_{0}}} {\partial
\Psi\ \ }\ \ } \ =\  \Psi
\  \frac{\partial \ln{f_{1}(\beta, \nu, \lambda, x_{0})}}
{\partial \beta\ \ \ \ \ \ \ \ \ \ \ \ \ \ \ \ \ \ \ },
\end{equation} 
where we have used Eqs.(2.23) and (2.24). Eq.(4.2) can also be obtained
from Eq.(4.1). Now introducing Eq.(4.2) in Eq.(2.13), we get  $\frac{d
{\tilde{\mathcal{N}}}} {d \beta\ \ }  =  0$. In other words, permissible
values of $\beta$ are determined from the extrema of $\tilde{\mathcal{N}}$,
as a function of $\beta$. The determination of extrema in turn needs
Eq.(4.2), Eqs.(2.20) and (2.26). 

Before we proceed further, we observe the following. Here, we have a 
variation problem involving two variables, $\Psi$ and $\beta$. When $\Psi$
is expressed as a function of $\beta$, we obtain the Hamiltonian,
${\rm{\tilde{H}}}   =  {\rm{\tilde{H}}}(\beta)$, and SLSs are determined
from its extrema, which are obtained  by setting $\frac{d
{\rm{\tilde{H}}}} {d \beta}  =  0$. Of course, in stead of $\beta$, we
could have used $\Psi$ as the fundamental variable. Now, when
${\tilde{\mathcal{N}}}$ is constant, the structure of ${\rm{\tilde{H}}}$
(Eq.(2.7)) is such that its extrema are determined from the extrema of
${\rm{\tilde{H}}_{0}}$. We have already investigated here this part. On
the other hand, we also have the option to take ${\rm{\tilde{H}}_{0}}$
constant, as it is done in this
sections and in sections to follow. In this limit the extrema
of ${\rm{\tilde{H}}}$ are determined from the extrema of
${\tilde{\mathcal{N}}}$, provided ${\tilde{\mathcal{N}}}(\beta)$  has
extrema. Another equivalent way of envisioning the problem comes from
Eq.(2.13), which is, of course the direct consequence of the structure of
the Hamiltonian, ${\rm{\tilde{H}}}$ (Eq.(2.2)). We can think of an
effective dynamical system, having two conjugate dynamical variables,
$\beta$ and $\Psi$. Then, the Poisson bracket formula, Eq.(2.13) suggests
that the effective or the reduced dynamical system can  be described by
the Hamiltonian, ${\rm{\tilde{H}}_{0}}(\Psi, \beta)$ having a constant of
motion, ${\tilde{\mathcal{N}}}(\Psi, \beta)$, or vice versa. Stationary
localized states in this dynamical system  picture are determined by fixed
points (FPs) of the effective or the reduced dynamical system. The two
sets of extrema, obtained from two procedures or two pictures may not be
identical. So, in the following section, I investigate this aspect of the
problem.

\subsection{Equation for  the fixed points of the reduced dynamical
system and results}

Now, if we  altogether ignore the infinite sum in Eq.(2.25) which
defines ${\tilde{\mathcal{N}}}(\Psi, \beta, x_{0})$, we obtain

\begin{equation}
\sqrt{\Psi}\ =\ \sinh{[\frac{\beta} {\sqrt{\Psi (1 +\ \Psi)} }\ \frac{d
\Psi} {d \beta}]}. 
\end{equation}  
\noindent
Introducing Eqs.(4.1) and (4.2) in Eq.(4.3), we get the equation which
determines $\beta$. If we are mostly interested in the roots of
Eq.(4.3), having magnitude less than unit magnitude, we can as before
ignore altogether the infinite  sum in Eq.(2.21), which defines
$f_{1}(\beta, \nu, \lambda, x_{0})$. This further simplifies the equation,
which determines $\beta$. 

When $\nu  =  0$, using Eqs.(2.21) and (4.1) we find that $\beta_{s}  =
{\rm{arc}}\sinh{a}$ makes the Eq.(4.3) an identity. We shall later show
that even with the full expression of $\frac{d {\tilde{\mathcal{N}}}} {d
\beta\ \ }$ , derivable from Eq.(2.25), the above  choice of $\beta_{s}$
makes $\frac{d {\tilde{\mathcal{N}}}} {d \beta\ \ }$ identically zero. In
other words, irrespective of ${\tilde{\mathcal{N}}}$ or 
${\rm{\tilde{H}}_{0}}$ is taken to be a constant in this constrained
variational approach, we get the same stationary AL solitons with $\omega 
=  - 2 \lambda \cosh{\beta}$ in both cases. Furthermore, when $\nu
\sim  o(1)$, we expect from this result that $\beta_{s} \sim
{\rm{arc}}\sinh{a}$. This is also borne out in our numerical
calculation, albeit not shown here. We shall show it in the exact
calculation.

Considering Eq.(4.3), we consider first the unstaggered localized
states with $\nu  > 0$.  First of all, for every value of $a  > 0$, we
find a $\nu_{{\rm{critical}}}$, such that for $\nu >
\nu_{{\rm{critical}}}$, Eq.(4.3) has no real root. On the other hand, $\nu
< \nu_{{\rm{critical}}}(a)$, we find two roots of Eq.(4.3) for a given
value of $a >  0$. It is also found that $\beta_{s}$ is a monotonically
increasing function of $\nu$ while $\beta_{l}$ is a monotonically
decreasing function of $\nu$. These features of the solutions are  shown
in Fig.8. 
\begin{figure}
\parindent 0.3in
\includegraphics{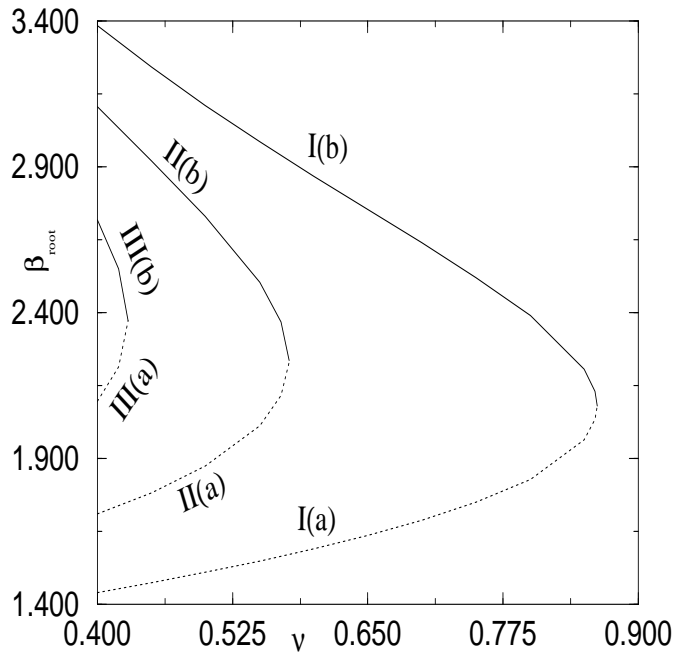}
\caption{This figure shows two real roots of Eq.(4.3) in the
text as a  function of the nonintegrability parameter, $\nu$ for three
values of the parameter, $a$. Since, $\lambda = 1 $, these states are
unstaggered stationary localized states. Curves I(a), I(b) : $a$ = 1.5,
Curves II(a) and II(b) : $a$ = 1.75, and Curves III(a) and III(b) $a$ =
2.0. While (a) or the lower part of all curves is for the smaller root,
$\beta_{s}$, (b) part or the upper part of these curves show the variation
of the larger root $\beta_{l}$. Note that this figure also shows the
variation of $\nu_{\rm{crtical}}$ as a function of the parameter, $a$.}
\end{figure}
Then, according to our previous discussion, stationary localized
states characterized by $\beta_{s}$ are stable, while the states
characterized by $\beta_{l}$ are unstable. In case of staggered localized
states  having $\nu > 0$, we find that for $|\nu \lambda|  >  1 $,
Eq.(4.3) has no root. Furthermore, $0  < |\nu \lambda| < 1$, Eq.(4.3) has
only one root, $\beta_{s}$. We find for $\lambda  = -1$, $\beta_{s}$
decreases with increasing $\nu$. This is shown in Fig.9. 
\begin{figure}
\parindent 0.35in
\includegraphics{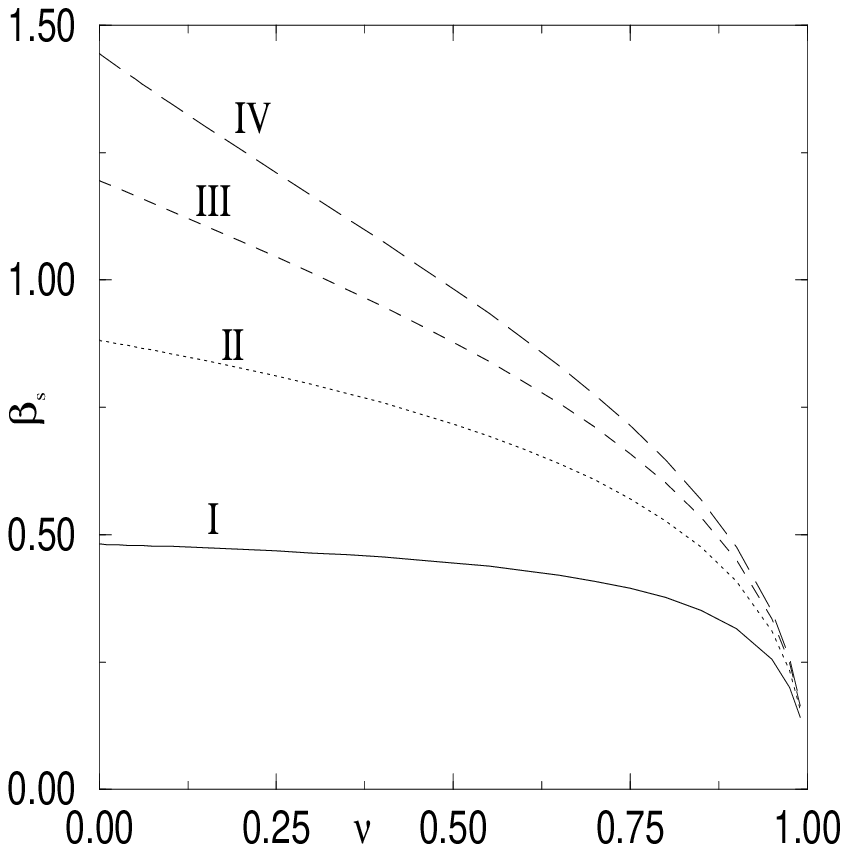}
\caption{This figure shows the variation of the smaller root,   
$\beta_{s}$ of Eq.(4.3) in the text as a function of the nonintegrability
parameter, $\nu$ for various values
of the parameter, $a$ . Since,
$\lambda  =  - 1 $, these states are staggered stationary localized 
states. Curve I : $a$ = 0.5, Curve II : $a$ = 1.0, Curve III : $a$ = 1.5,
and Curve IV : $a$ = 2.0.}
\parindent 0.35in
\includegraphics{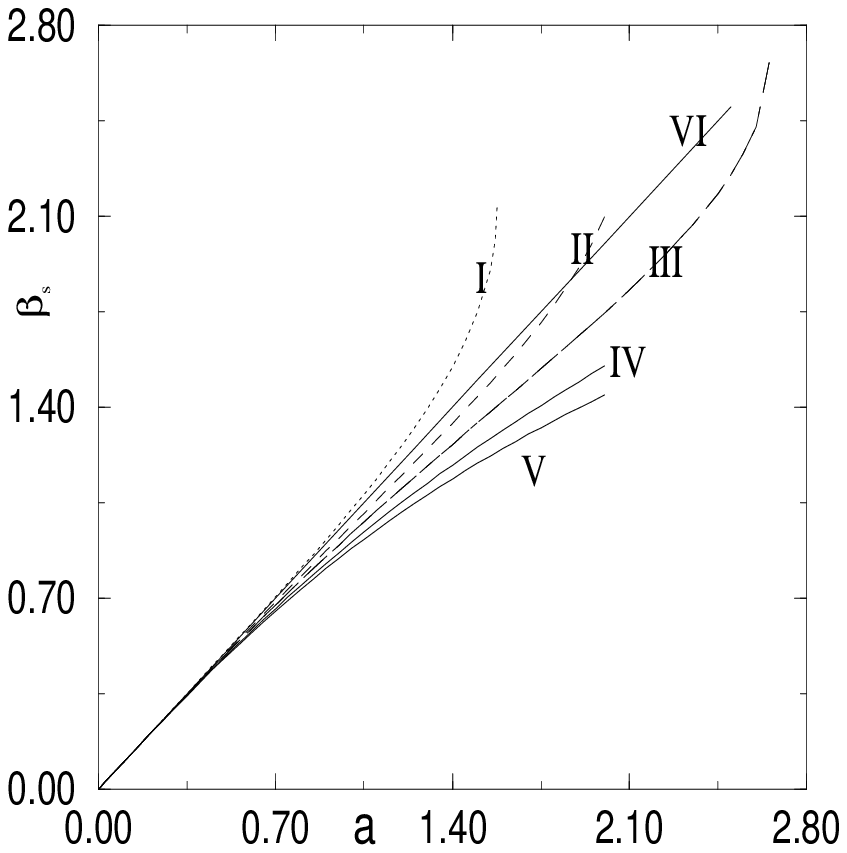}
\caption{This figure shows the variation of the smaller root,
$\beta_{s}$ of Eq.(4.3) in the text as a function of the parameter, $a$
for various values of the nonintegrability parameter, $\nu$. Since,
$\lambda = 1 $, these states are unstaggered stationary localized states.
Curve I : $\nu$ = 0.75, Curve II : $\nu$ = 0.40, Curve III : $\nu$ =
0.25, Curve IV : $\nu$ = 0.0. Curve VI is the straight line, $\beta_{s} =
a$.}
\end{figure}
Since, $\beta^{-1}_{s}$ gives a measure of the width of
the localized states, from our finding we conclude that staggered
stationary localized states vanish whenever $|\nu \lambda| \ge 1$. 
This happens due to effectively repulsive on-site nonlinear potential.
This potential helps spread the amplitude over the whole sample. We have
already mentioned this.  It is also found for both unstaggered
and staggered cases that for $a \sim  o(1), \ \ a  \sim  \sinh{\beta_{s}}
\sim  \beta_{s}$, and hence, in this asymptotic limit, Eq.(4.1) together
with Eq.(2.21) yields $\sqrt{\Psi}\ \sim \frac{\beta_{s}} {\sqrt{1.0\ +\
\nu \lambda}}.$ Fig.10 shows the result for the unstaggered states with
$\nu > 0$. However, the result for staggered localized states is not shown
here. This result agrees with the asymptotic result in Ref.(32).

The next important aspect is to study the effect of $x_{0}$  on the 
formation of these states. Another equally important aspect is to examine
if unstable localized states that we find in the truncated equations
or equivalently in the leading term analysis exist
in the exact calculation. We first emphasize that the problem can be
solved exactly in this reformulated version. For our numerical analysis,
we use the "FindMinimum" program of "MATHEMATICA -version 4". We discuss
below the exact solution.

\subsection{The exact solution}

We consider first the case of unstaggered stationary localized states for
$\nu > 0$ and $x_{0}  = 0.0.$  Fig.11 shows the variation of
$\beta_{\rm{root}}$ as a function of the parameter, $a$ for various
values of the nonintegrability parameter, $\nu$. The corresponding figure
to be compared is Fig.10. 
\begin{figure}
\parindent 0.35in
\includegraphics{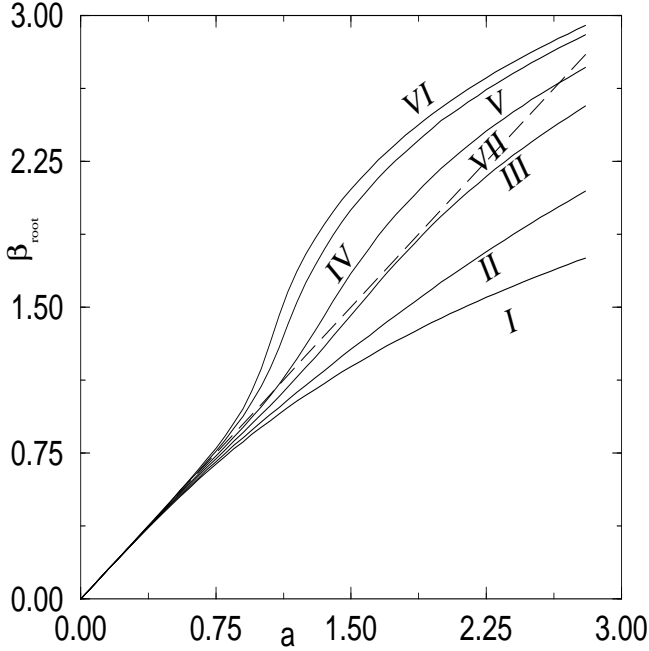}
\caption{This figure shows the variation of $\beta_{\rm{root}}$ as a
function of the parameter, $a$, for various values of the nonintegrability
parameter, $\nu$. This is obtained  from the exact calculation. For this
figure, ${\tilde{\rm{H}}_{0}} = {\rm{constant}},\  x_{0} = 0.0,\   
{\rm{and}}\ \lambda = 1$. Curve I : $\nu$ = 0.0, Curve II : $\nu$ = 0.10,
Curve III : $\nu$ = 0.40, Curve IV : $\nu$ = 0.75, Curve V : $\nu$ = 0.90,
and Curve VI : $\nu$ = 1.0.  Curve VII, the dashed curve  is the straight
line, $\beta_{\rm{root}} = a$.}
\end{figure}
We note that for $\nu = 0.0$, we obtain from Eq.(4.1) and Eq.(2.21),
$\Psi(\beta, 0.0, \lambda, x_{0})  =  a \sinh{\beta}$. On the other hand,
analytically $\Psi(\beta, 0.0, \lambda, x_{0})  =  \sinh^{2}{\beta}$.  So,
we must have then $a  = \sinh{\beta_{\rm{root}}}$. This is clearly obtained 
in our numerical analysis. As $\nu = 0.0 $ for these curves, both the
Curve I of Fig.11 and the Curve V of Fig.10 are defined by the equation
$\beta_{\rm{root}} = {\rm{arc}}\sinh{a}$. We further see in Fig.11 and
also in Fig.10 that for small values of $a$ all curves merge
simultaneously on the line $ \beta_{\rm{root}} = a$ and Curve I
(Fig.11) or Curve V (Fig10). This in turn implies that for $a \sim  o(1)$,
$a \sim \sinh{\beta_{\rm{root}}} \sim \beta_{\rm{root}}$.  Hence, for
on-site peaked unstaggered localized states, $\sqrt{\Psi}\ \sim 
\frac{\beta_{\rm{root}}} {\sqrt{1.0\ +\ \nu}}$ is the asymptotic
result\cite{32}. We further note that in the exact calculation, we do not
find any root corresponding to $\beta_{l}$ of Eq.(4.3) for any value of
$a$. This conclusion is reached from the following observation
in our numerical analysis. In our numerical analysis, we have used N and 
M number of terms in two infinite sums in Eqs.(2.21) and
(2.25) respectively. We find that $\beta_{l} \rightarrow \infty$
monotonically if both $ N\ {\rm{and}}\ M \rightarrow \infty$, either
separately or simultaneously. So, unstable stationary localized states
obtained from Eq.(4.3) are spurious and due to the truncation
error. Similar argument should hold good for the analysis of Eq.(3.11).

The variation of $\beta_{\rm{root}}$ of unstaggered localized states as a
function of $\nu$ for various values of $a$ from the exact calculation,
and this is shown in Fig.12. 
\begin{figure}
\parindent 0.25in
\includegraphics{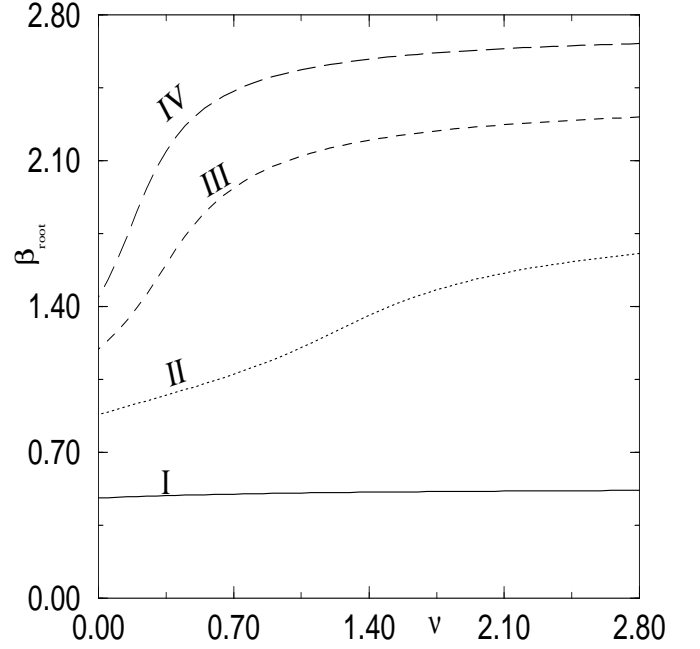}
\caption{This figure shows the variation of $\beta_{\rm{root}}$ as a
function of the nonintegrability parameter, $\nu$  for various values of
the parameter, $a$. It is obtained  from the exact calculation. For
this figure, ${\tilde{\rm{H}}_{0}} = {\rm{constant}},\  x_{0} = 0.0,\
{\rm{and}}\ \lambda = 1$. Curve I : $a$ = 0.5, Curve II : $a$ = 1.0, Curve
III : $a$ = 1.5, and Curve IV : $a$ = 2.0.}
\end{figure}
We see that $\beta_{\rm{root}}$ is a monotonically increasing function of
$\nu$ for $\nu > 0.0$. We note that $\beta^{-1}_{{\rm{root}}}$ gives a
measure of the width of the localized states. We have also argued before
that the width of the stable unstaggered localized states for this case
must decrease with increasing $\nu$. So, $|\beta_{{\rm{root}}}|$ must
increase with increasing $\nu$, if it were to characterize stable
SLSs. Inasmuch as $\beta_{\rm{root}}$ satisfies this criterion, stationary
localized states corresponding to $\beta_{\rm{root}}$ are stable.

For inter-site peaked unstaggered localized states, having  $x_{0}\ =\
0.5 $, the dependence of $\beta_{\rm{root}}$ on $a$ for a fixed $\nu$ is
also investigated for various values of $\nu > 0$. It is shown in Fig.13.
Fig.13 also includes for comparison the variation of $\beta_{\rm{root}}$
as a function of $a$ for $x_{0}\ =\ 0.0 $. 
\begin{figure}
\parindent 0.3in
\includegraphics{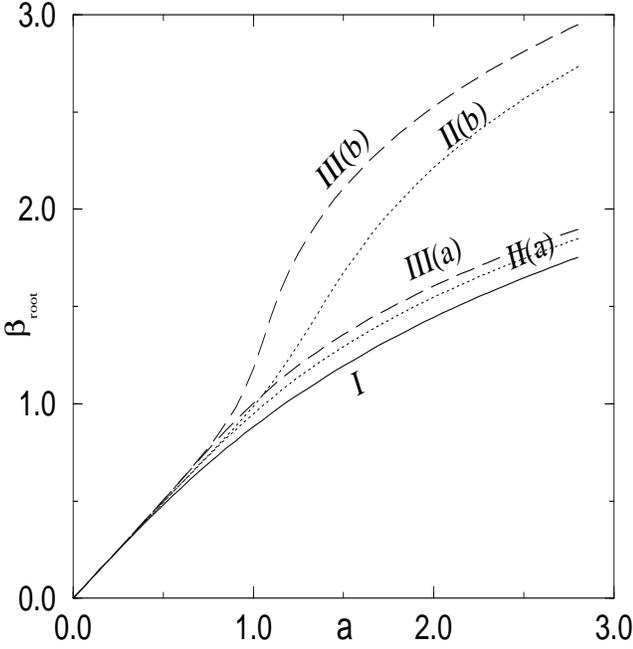}
\caption{This figure shows the variation of $\beta_{\rm{root}}$  as a
function of the parameter, $a$ for two positive values of $\nu$. This   
figure presents the exact solution for $\lambda = 1$. For the Curve I,
$\nu = 0.0$. For Curves II(a) and II(b) $\nu = 0.40$, but $x_{0} = 0.5\
{\rm{and}}\ 0.0$ respectively. For Curves III(a) and III(b) $\nu = 1.0$,  
but $x_{0} = 0.5\ {\rm{and}}\ 0.0$ respectively.}
\end{figure}
We note that localized states have larger widths for $x_{0}\ =\
0.5$. This, in turn implies that inter-site peaked SLSs will have higher
energy. Similarly, the variation of $\beta_{\rm{root}}$ as a function of
$\nu$ for $x_{0}\ =\ 0.5 $, is shown in Fig.14. We again note that
stationary localized states for $x_{0}\ =\ 0.5$ have larger widths. Most
importantly inter-site peaked SLSs show weak dependence on $\nu$. 
\begin{figure}
\parindent 0.35in
\includegraphics{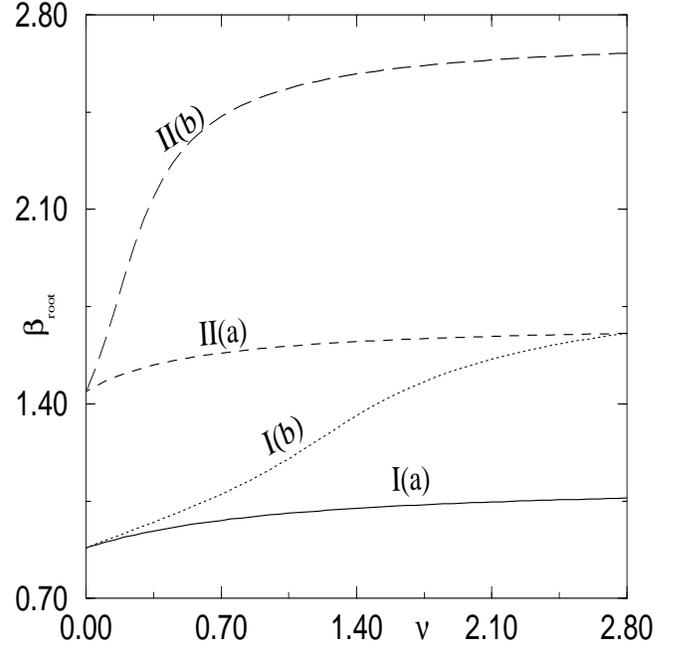}
\caption{This figure shows the variation of $\beta_{\rm{root}}$  as a
function of the nonintegrability parameter, $\nu$ for two positive values
of the parameter, $a$. This figure presents the exact solution for
$\lambda = 1$. For Curves I(a) and I(b) $a = 1$, but $x_{0} = 0.5\
{\rm{and}}\ 0.0$ respectively. For Curves II(a) and II(b) $a = 2.0$,
but $x_{0} = 0.5\ {\rm{and}}\ 0.0$ respectively.}
\end{figure}

In Fig.15  we show the variation of energy of the on-site and inter-site
peaked stationary localized states as a function of $\nu$ for $x_{0}\ =
0.0\   {\rm{and}}\  0.5$. When $\nu  \sim  o(1)$, the solution
approximately has the continuous symmetry of the solution of $\nu  =
0.0$\cite{24, 25}. So, in this limit, both on-site and inter-site peaked
states should have almost the same energy. This is clearly seen in
Fig.15. 
\begin{figure}
\parindent 0.15in
\includegraphics{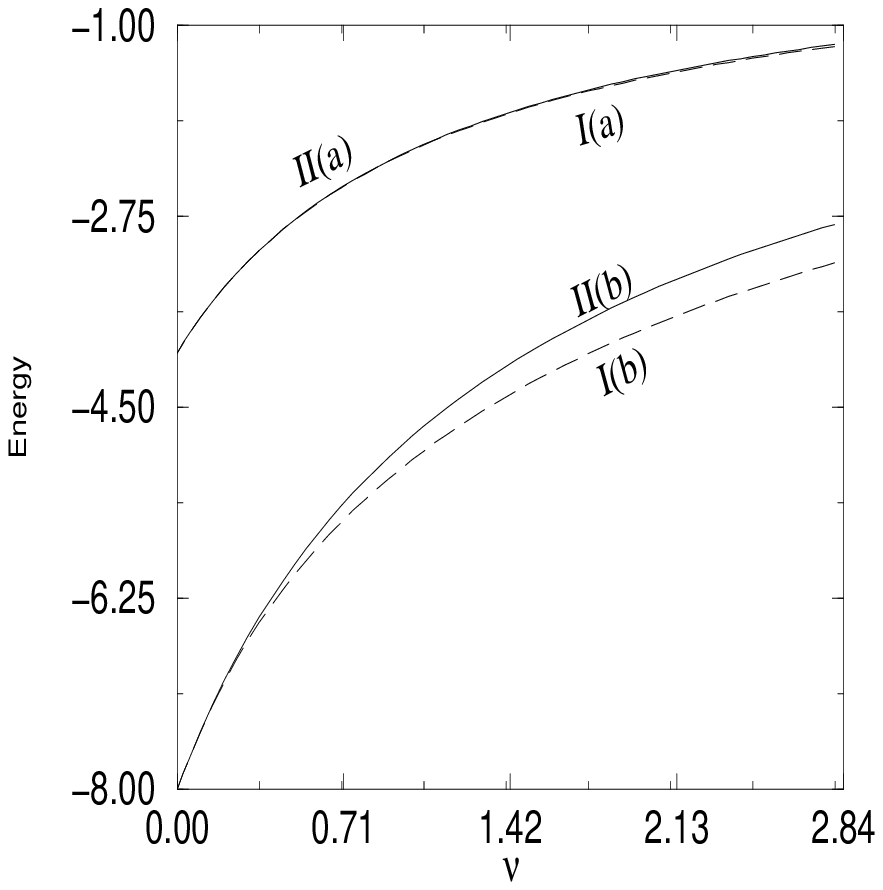}
\caption{This figure shows the variation of the energy of
unstaggered stationary localized states as a function of $\nu$ for two   
values of $a$ and for two permissible values of $x_{0}$. Of course, the
result is obtained from the exact calculation with  ${\tilde{\rm{H}}_{0}}
= {\rm{constant}}$.  For Curves I(a) and II(a) $a = 1.0$, but $x_{0} =
0.0\ {\rm{and}}\ 0.5$ respectively. For Curves I(b) and II(b) $a = 2.0$,
but $x_{0} = 0.0\ {\rm{and}}\ 0.5$ respectively.}
\end{figure}
Similarly, on-site peaked states are supposed to be more stable
than inter-site peaked states. This is also clearly seen by comparing
Curves I(b) and II(b) in this figure. Again, when $a$ reduces, the width
of the corresponding stationary localized state increases. Consequently,
the energetic distinction between the on-site and inter-site peaked states
reduces. This is also clearly evident in Fig.15.

A comprehensive understanding of these results, delineating the basic
differences of on-site peaked and inter-site peaked unstaggered SLSs is
definitely required. To this end, we note that for $\nu > 0$, these
results indicate the operation of a nonlinear attractive potential in the
system. This effective nonlinear potential is maximally attractive at
lattice sites. From the physical consideration, we argue that the
attractive potential assumes minimum values at the center of any two
consecutive lattice sites. Since, the system has lattice translational
invariance, this potential will also have the periodicity of the
underlying lattice.  An attractive potential effectively reduces the
inter-site hopping of a
particle, consequently helping its localization.  Secondly, any state can
be thought of an effective particle with an effective mass, executing a
motion in a potential. From the physical consideration, it is easy to see
that stronger is an attractive potential, stronger is the
localization. Consequently, heavier is the effective mass of the
particle. Conversely then the inverse of the effective mass gives the
localization length of the effective state. In this picture then the
unstaggered SLS for $\nu > 0 $ is equivalent to an effective particle
sitting either at the bottom of any well ($x_{0} = 0.0$) or at the top of
the same well ($x_{0} = 0.5$). So, the unstaggered SLS with $x_{0} = 0.0$
will corresponds to a heavier effective mass particle than the
corresponding SLS with $x_{0} = 0.5$. In terms of localization length, the
first kind of states will be more localized than the second type. Another
important deduction from this picture is that energetically, the first
kind of states should be more stable. These results are seen in our
numerical analysis. Again, when we increase the parameter, $a$, we
increase the maximum amplitude of the SLS. In this effective picture, the
depth of the potential well increases. Similar situation also occurs by
increasing $\nu$. This, in turn implies that the effective mass of the
particle at the bottom of the well will increase with increasing $a$ and
$\nu$. So, in both cases, the width of the on-site peaked unstaggered SLS
should decrease, as seen in our numerical calculation. The effective
periodic potential, however is a function of at least three variables, the
position variable, $x_{0}$, the parameter, $a$ and the nonintegrability
parameter, $\nu$. Since, the SLS with $x_{0} = 0.5$ shows weak dependence
on $\nu$, our results suggest that the top of the potential is not
substantially affected by the change in $\nu$. On the other hand, from our
results we deduce that the top of the potential is energetically
stabilized  by increase in $a$.
 
Finally, some exact calculations of $\beta_{\rm{root}}$ for staggered SLSs
for $\nu  >  0,\ \lambda  =  -1$  and for both $x_{0} =  0.0 \ {\rm{and}}\
0.5$ are presented. In Fig.16 we present the variation of $\beta_{\rm{root}}$ 
as a function of $\nu$ for $a = 1.0 \ {\rm{and}}\ 2.0 $. 
\begin{figure}
\parindent 0.3in
\includegraphics{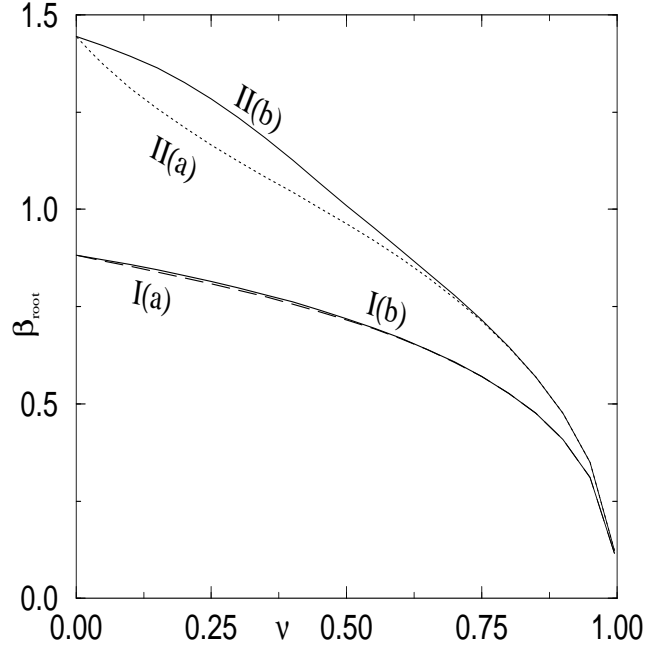}
\caption{This figure shows the variation of $\beta_{{\rm{root}}}$ as   
a function of the nonintegrability parameter, $\nu$ for $a  = 1.0 \
{\rm{and}}\ 2.0$, as obtained from the exact calculation.  Since,
$\lambda = - 1$, these states are staggered stationary localized
states. Curve I(a) : $a$ = 1.0, and $x_{0} = 0.0$. Curve II(a) : $a$ =  
2.0, and $x_{0} = 0.0$.  Curve I(b) : $a$ = 1.0, and $x_{0} = 0.5$.  Curve
II(b) : $a$ = 2.0, and $x_{0} = 0.5$.}
\end{figure}
Consider first $x_{0} = 0.0.$  Curves I(a) and II(a) in Fig.16 are almost
identical to corresponding curves Fig.9 with a discernible deviation in
the magnitude of $\beta_{\rm{root}}$ for large values of $a$ together with
small values of $\nu$. The same calculation with $\tilde{\mathcal{N}}\ =\
{\rm{constant}}$ gives $\beta_{\rm{s}} \rightarrow  0 $ linearly as $ \nu
\lambda \rightarrow -1$. See Fig.5. Fig.17 shows the variation of
$\beta_{\rm{root}}$ as a function of $a$ for various values of $\nu >
0$. 
\begin{figure}
\parindent 0.3in
\includegraphics{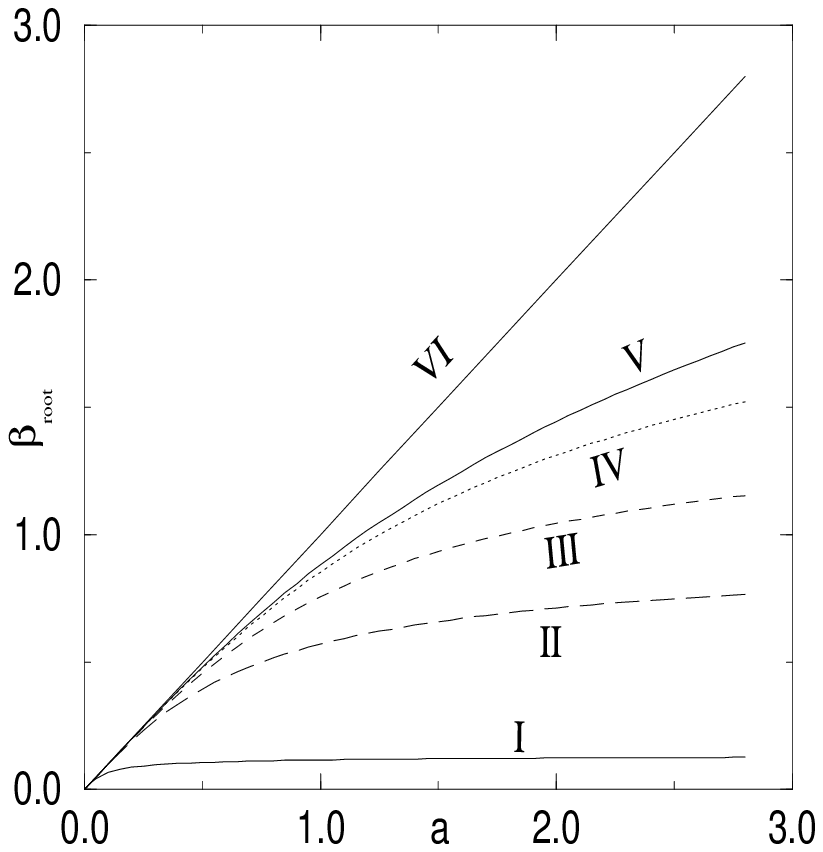}
\caption{This figure shows the variation of $\beta_{\rm{root}}$  as a  
function of the parameter, $a$ for  positive values of $\nu$. This 
figure presents the exact solution for $\lambda = - 1$ and $x_{0} =
0.0$ . Curve I : $\nu$ = 0.995, Curve II : $\nu$ = 0.75, Curve III :
$\nu$ = 0.40, Curve IV : $\nu$ = 0.1, and  Curve V : $\nu$ = 0.0. Curve 
VI, the solid curve is the straight line, $ \beta_{\rm{root}} = a$.}
\end{figure}
The important point to note is that $\beta_{\rm{root}}(\nu \lambda <
0) < \beta_{root}(\nu = 0 )$. In other words, SLSs for $\nu > 0$ are
stabilized by expansion of the width. As $\beta_{\rm{root}} = a$
line becomes tangent to all curves in Fig.17, we infer that when $a \sim
o(1)$, $ \beta_{\rm{root}} \rightarrow a$. Again on the
curve $\nu = 0$, $ \sinh{\beta_{\rm{root}}} = a $.  In this asymptotic
limit, we then have from Eqs.(2.21) and (4.1) that $\sqrt{\Psi} \sim 
\frac{\beta_{\rm{root}}} {\sqrt{1 - \nu}}$. The corresponding approximate
calculation for the model with  $\tilde{\mathcal{N}}\ =\ {\rm{constant}}$
is shown Fig.6.  Fig.6 shows that when $\alpha \sim o(1)$, $\beta_{\rm{s}}
\sim \alpha^{2}(1 - \nu)$. This in turn yields the same asymptotic result
for $\sqrt{\Psi}$. Comparing our results for staggered SLSs from three
different approaches, namely, approximate calculations with
(a) $\tilde{\mathcal{N}}\ =\ {\rm{constant}}$, (b) $ {\rm{\tilde{H}}_{0}}\
=\ {\rm{constant}}$, and (c) the exact calculation with
${\rm{\tilde{H}}_{0}}\ =\ {\rm{constant}}$ and $x_{0} = 0.0$, we conclude
that all three give qualitatively the same result.

We now consider the basic difference between staggered SLS with $x_{0} =
0.0$ and staggered SLS with $x_{0} = 0.5$. Fig.16 shows the the variation
of $\beta_{\rm{root}}$ as a function of $\nu \ge 0$ for  $a = 1.0 \
{\rm{and}}\ 2.0.$ From this figure, we see that though $\beta_{\rm{root}}
\rightarrow 0 $ as $\nu \rightarrow 1$ for both $x_{0} = 0.0, \
{\rm{and}}\ 0.5$, the magnitude of $\beta_{\rm{root}}$ for intermediate
values of $\nu$ is greater for the SLS with $x_{0} = 0.5$. This, in turn
implies that SLS with $x_{0} = 0.5$ has smaller localization length. This
is very much opposite to what we observe for an unstaggered SLS. 
We consider next Fig.18. 
\begin{figure}
\parindent 0.3in
\includegraphics{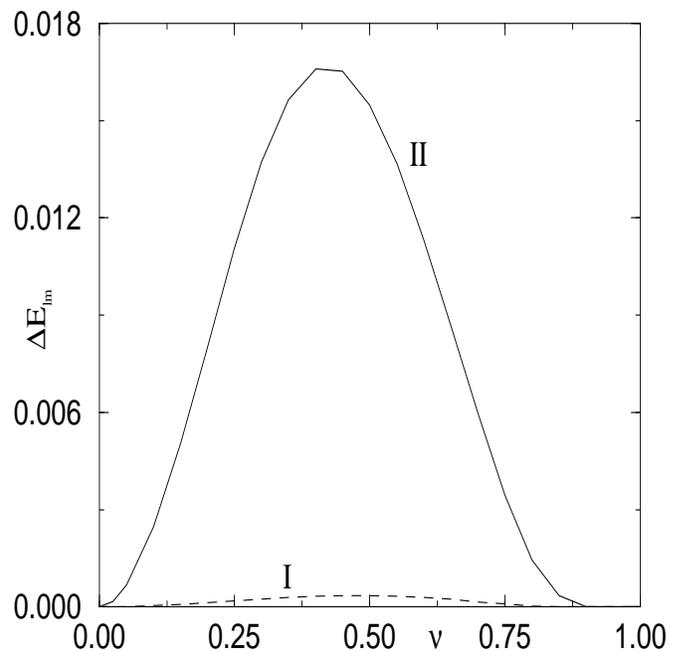}
\caption{$\Delta E_{lm} = E_{l} - E_{m}$, where
$E_{l}$ and $E_{m}$ define the energy of staggered SLS with $x_{0} = 0.0$
and $x_{0} = 0.5$ respectively. This figure  shows the variation of
$\Delta E_{lm}$ as a function for $\nu \in [0. 1).$ Curve I : $a$ = 1.0,
and Curve II : $a$ = 2.0.}
\end{figure}
We define $\Delta E_{lm} = E_{l} - E_{m}$, where
$E_{l}$ and $E_{m}$ define the energy of staggered SLS with $x_{0} = 0.0$
and $x_{0} = 0.5$ respectively. Fig.18 shows the variation of
$\Delta E_{lm}$ as a function for $\nu \in [0. 1).$  As $\Delta E_{lm} >
0$ for intermediate values of $\nu$, SLS with $x_{0} = 0.5$ is
energetically stable compared to its equal counterpart.

In further analysis, we note that the amplitude distribution of staggered 
SLS around the maximum amplitude site is ST mode like for $x_{0} = 0.0$
and P mode like for $x_{0} = 0.5$. In this case also there is a
periodic arrangement of potential wells in the system. Since, the
on-site nonlinear potential is repulsive in this case, this
potential will attain the minimum at the center of two consecutive 
lattice sites. So, the whole periodic arrangement of wells is shifted by
half a lattice constant. Consequently, the effective particles
corresponding to the P like mode and the ST
like mode are sitting at the bottom of a well and at the top of a well
respectively\cite{50}. So, the P like mode corresponds to an effective
particle with larger effective mass than the corresponding ST like mode,
and thereby having a smaller localization length. From this picture, we
also deduce that the P like mode is energetically more stable than the ST
like mode. Most importantly however, we prove by our variational approach
that the existence of P like mode and ST like mode is a fundamental
property of a system described by IN-DNLS. To successfully explain
the dependence of the localization length of the P like modes
on these two parameters, $a$ and $\nu$ respectively, we need the following
behavior of the potential. When the parameter, $a$ increases, the
effective well depth must increase  for intermediate values of $\nu$.  
Consequently, the effective mass of the particle will increase, and
the localization length will decrease.  But, in case of $\nu$, the
effective well depth must decrease as $\nu \rightarrow 1$. Furthermore, 
as discussed in the context of unstaggered SLS, the dependence of the
localization length of ST like modes determines the dependence of the
top of the potential well on two important parameters, $a$ and $\nu$.

\section{summary}
IN-DNLS is a one dimensional discrete nonlinear equation with a tunable
nonintegrability parameter, $\nu$\cite{32,33}. When $\nu = 0$, it reduces
to the famous AL equations\cite{24,25}. The importance of IN-DNLS in
physics as well as in nonlinear mathematics is discussed in the text. In
this paper, primarily eigenvalues, energies and corresponding
site-amplitudes of SLSs of IN-DNLS are studied using discrete variational
formulation\cite{23, 42, 52}. The standard variational approach starts
from the respective Lagrangian to study this type of problem. In this
paper, however the appropriate functional is derived using the standard
variational procedure for finding eigenvalues of Sturm-Liouville
equations\cite{58}. In other words, it is shown here how the effective
functional can be derived from the Hamiltonian and constants of motion
without the prior knowledge of the Lagrangian. The uniqueness of the
functional is also established by showing its equivalence to the effective
Lagrangian.    

Inasmuch as localized states in one dimensional linear impure as 
well as disordered systems show asymptotic exponential decay\cite{37, 38},
a "sech'" ansatz with two parameters, $\beta$ and $\Phi = \sqrt{\Psi}$ is
used to find eigenvalues and eigenfunctions. In this choice, $\beta^{-1}$
and $\sqrt{\Psi}$ define the width and the maximum amplitude of SLS
respectively. This ansatz is so chosen as it gives AL stationary localized
states when $\nu \rightarrow 0$. Furthermore, SLSs of IN-DNLS are assumed
to belong to the class of breathers with a single
frequency\cite{12}. Since, stationary solitons of AL equations are
breathers of this class, this choice of form for SLSs of IN-DNLS is
justified.    

Very naturally two procedures have emanated in our variational
calculation. In the first case, the reduced dynamical system is described
by the Hamiltonian, ${{\tilde{H}}_{0}}$ and  $\tilde{\mathcal{N}}$ is
taken to be the number constant. In the second case, $\tilde{\mathcal{N}}$
acts as the Hamiltonian. Since, the analysis involves two infinite sums,
both sums are ignored in both cases in the leading term analysis. In both
cases, for unstaggered stationary localized states two permissible values,
namely $\beta_{s}$ and $\beta_{l}$ of the width parameter, $\beta$, are
found. It is further found that for two real roots to exist we need $\nu <
\nu_{\rm{critical}}$  if $\nu > 0$.  Furthermore, $\nu_{\rm{critical}}$
is found to be a monotonically decreasing function of the parameter,
$\alpha$, in the first case and $a$ in the second case. These parameters
are  defined in the text and are positive semidefinite. It is  
successfully argued from our numerical analysis that SLSs,
characterized by the smaller width parameter, $\beta_{s}$ are stable and
states characterized by $\beta_{l}$ are unstable. However, for staggered
SLSs both procedures have yielded a single value, $\beta_{s}$ and our
numerical results indicate that these are stable localized modes of the
system. Though both procedures yield qualitatively same results for
the parameters of SLSs, no quantitative comparison is attempted
here. Again, it is found that the problem can be exactly solved in the
second case. In our exact solution no unstable SLS is obtained. So, the
occurrence of unstable SLSs in this system in the leading term analysis 
should be attributed to the truncation error. In the context of SLS in one
dimensional nonlinear systems, this is indeed an important result.

The formation of unstaggered and staggered SLSs are investigated here for
$\nu \ge 0$. For the null value of $\nu$, the present variational
procedure correctly produces SLSs of AL equation. Furthermore, when $\nu
\rightarrow 0$, it is found that $\beta_{s} \rightarrow \alpha^{2}$ in the
first case for both unstaggered and staggered SLSs. Consequently, AL
stationary soliton is recovered in this asymptotic limit. In the second
case, $a \rightarrow \sinh{\beta_{s}}$ asymptotically as $\nu \rightarrow
0$ in the leading term analysis. This result is true for both unstaggered
and staggered SLSs. The same asymptotic results are found in the exact
analysis too, except that $\beta_{s}$ is replaced by $\beta_{\rm{root}}$.   
So, in the second case too  the AL stationary soliton is the asymptotic
result for $\nu \rightarrow 0$. Analytically also the same asymptotic
result is obtained here. Our both analytical and numerical results are
expected on physical consideration.

In the other asymptotic analysis, $\alpha \rightarrow 0$ in the first case
and $a \rightarrow 0$ in the second case. Again, in the second case there
are two scenarios, the leading term analysis and the exact calculation.
For all cases and for both unstaggered and staggered localized states,
the known asymptotic form of $\Psi$ for the stable SLS are obtained
numerically from the present variational analysis. This is, therefore a
very important contribution of the present work. For unstaggered SLSs, it
is found that the width of the state decreases with increasing $\nu$. On
the other hand, for staggered SLSs, the width increases with increasing
$\nu$ and vanishes as $\nu \rightarrow 1$. These results are consistent
with the physics of the problem and reasons are given in the text.  

Another important aspect is the dependence of the width of SLSs on the
position of the maximum amplitude, denoted by $x_{0}$. It is proved in
the text that $x_{0} = 0\ {\rm{or}}\  \pm \frac{1} {2}$.
It is observed in our analysis that for unstaggered SLSs the on-site
peaked SLS ($x_{0} = 0$) has smaller width than the inter-site peaked
($x_{0} = \pm \frac{1} {2}$) SLS for a given value $\nu > 0$ and $a >
0$. Our analysis also shows that for a given $\nu$ and $a$, the on-site
peaked unstaggered SLS is energetically more stable than the corresponding
inter-site peaked SLS. These results are physically realistic and are
successfully explained using the effective mass picture. It is found in
our analysis that the existence of the P like mode and the ST like mode
is a fundamental property of the system, described by IN-DNLS. It is also
shown in this numeroanalytical method that the P like mode is
energetically more stable than the corresponding ST like mode. These
results constitute a very important contribution of the present work.

It is definitely important to find exact eigenvalues and eigenvectors of
the problem. Present analysis may turn out to be a useful guide for the
exact calculation. In this analysis we have not proved that the lowest
eigenvalue is obtained. Furthermore, the system may have more than one
SLS type of ILM. These questions need to be properly investigated. 
Presence of impurity in the nonintegrability parameter,
$\nu$ may produce more stationary localized states and these states 
may interact through further external perturbation. A study of this type
is important in the transport in nonlinear systems\cite{19,21}. In our
calculation, we find both  P like mode and  ST like mode. It will
be interesting to know the asymptotic form of these modes in this
model. Furthermore, the behavior of these modes under external
perturbation should also be investigated. Finally, it is important to find
more physical as well as biological systems, where IN-DNLS can be used to
study transport properties. A good candidate in this regard is the
transport across biological membranes of protons through proton-wires.

{\appendix
\section{The formulation of discrete variational approach for IN-DNLS}
We are dealing with a nonlinear eigenvalue problem in our aim to find
stationary localized states of IN-DNLS equation, Eqs.(2.2) and (2.5) in
the text. For this purpose we are employing variational
formulation\cite{23, 53, 58}. To implement the variational approach for
this problem, we require the proper functional, $\tilde{\rm{F}}$ whose
constrained variation will lead to Eq.(2.5)\cite{58}. We, of course know a
constant of motion and the Hamiltonian,  $\tilde{\mathcal{N}},\
{\rm{and}}\ \tilde{\rm{H}}$ respectively for the problem at hand\cite{32}.

Inasmuch as we know $\tilde{\mathcal{N}},\ {\rm{and}}\ \tilde{\rm{H}}$,
using the analogous variational approach of finding eigenvalues in
standard Sturm-Liouville problems\cite{ 58}, we set up the functional, 
$\tilde{\rm{F}}\ =\ \tilde{\rm{H}}\ -\ \Lambda\ \tilde{\mathcal{N}}$
where $\Lambda$ is the Lagrange multiplier\cite{59}. We then have for the
variation
of $\tilde{\rm{ F}}$, $\delta {\tilde{\rm{F}}}\ =\ \delta 
{\tilde{\rm{H}}}\ -\ \Lambda\ \delta {\tilde{\mathcal{N}}}$.
For the calculation of the variation, we transform $\Psi_{n}\ \rightarrow\
\Psi_{n}\ +\ \delta \Psi_{n},\ n\ \in Z$ in the expression of   
${\tilde {\rm{H}}}\ {\rm{and}}\ {\tilde{\mathcal{N}}}$ (Eqs.(2.6) and
(2.7) respectively) to obtain
\begin{widetext}
\begin{eqnarray}
\delta\ \tilde{\rm{F}}\ &=&\ \delta\ \tilde{\rm{H}}\ -\ \Lambda\
\delta\ \tilde{\mathcal{N}}\nonumber\\
&=& - 2 \sum_{n} \frac{\lambda\ (1 + \Psi_{n}^{2}) (\Psi_{n+1}\ +\
\Psi_{n-1})\ +\ 2\
\nu\ \Psi_{n}^{3}\ +\ \Lambda\ \Psi_{n}} {1 +\ \Psi_{n}^{2}}\ \delta
\Psi_{n}.
\end{eqnarray}
\end{widetext}
Since, $\{\delta \Psi_{n}\}$ are arbitrary, $\delta {\tilde{\rm{F}}}\ =\
0$ implies that\begin{equation}
 \lambda\ (1 + \Psi_{n}^{2}) (\Psi_{n+1}\ +\
\Psi_{n-1})\ +\ 2\
\nu\ \Psi_{n}^{3}\ +\ \Lambda\ \Psi_{n}\ =\ 0.
\end{equation}  
We note that (A.2) is identical to Eq.(2.5) when $\Lambda\ =\ \omega$.
From further analysis, we find that $\omega$ is given by Eq.(2.12) in the
text. 

For the case, where ${\rm{\tilde{H}_{0}}}\ =\ {\rm{constant}}$, the
corresponding
functional, ${\rm{\tilde{F}}}$ should be  given by  
${\rm{\tilde{F}}}\ =\ \Lambda_{2}\ {\rm{\tilde{H}_{0}}}\ +\
2\ \nu\ \tilde{\mathcal{N}}$, where $(\Lambda_{2}\ -\ 1)$ is the Lagrange
multiplier. The same procedure will yield Eq.(2.5) if $\Lambda_{2}\ =\
\frac{2 \nu} {2 \nu - \omega}$. 

So, it is important to note that we can devise the required functional to
determine the eigenvalues of SLSs by variational approach without the
formal knowledge of the Lagrangian.

\section{ Calculation of the function,
${\tilde{\mathcal{N}}}(\Psi, \beta, x_{0})$, Eq.(2.25)}

Before we proceed in this section, we cite some results required for the
calculation\cite{61}.
\begin{widetext}
\begin{equation}
I(s, \Psi, \beta))\ =\ \int_{0}^{\infty} dy\  \frac{\cos{{\frac{\pi s}
{\beta}} y}}
{\cosh{y}\ +\
(1\ +\ 2 \Psi)}\ =\ \frac{\pi\ \sin{\{\frac{2 \pi s} {\beta}} 
\ {\rm{arc}}\sinh{\sqrt{\Psi}}\}} {2\  \sqrt{\Psi\ (1\ +\ \Psi)}\
\sinh{\frac{\pi^{2} s} {\beta}}}. 
\end{equation}
\noindent
From (B1) we get 
\begin{equation}
\lim_{s \rightarrow\ 0}\ I(s, \Psi, \beta) = 
\frac{{\rm{arc}}\sinh{\sqrt{\Psi}}}
{\sqrt{\Psi\ (1\ +\ \Psi)}}\ =\  \frac{d\
({\rm{arc}}\sinh{\sqrt{\Psi}})^{2}} {d\ \Psi\ \ \ \ \ \ \ \
\ \ \ \ \ \ \ \ \ }.
\end{equation}
\end{widetext}
In our calculation, we are using the following ansatz.
\begin{equation}
\Psi_{n}\ =\ \Phi \frac{1} {\cosh\beta (n\ -\ x_{0})}\ ,\ n\ \in Z. 
\end{equation}
\noindent
This ansatz has also been used in the previous analysis\cite{32}. For
on-site peaked and ST like localized states, $x_{0}\ =\ 0$, and for
inter-site peaked and P like states, $x_{0}\ =\ \pm \frac{1}
{2}$\cite{23, 45, 46, 47}. We further write $\Phi^{2}\ =\ \Psi$.
The function, ${\tilde{\mathcal{N}}}(\Psi, \beta, x_{0})$ is given by
Eq.(2.6) in the text. Now introducing (B3) in Eq.(2.6) and then taking
partial derivative with respect to $\Psi$, we get

\begin{equation} 
\frac{\partial \tilde{\mathcal{N}}} {\partial \Psi\ \ }\ =\ \sum_{n\ =
- \infty}^{\infty} \frac{1} {\cosh^{2}{\beta (n\ -\ x_{0})}\ +\ \Psi}   
\end{equation}
\noindent
We use next the famous Poisson's sum formula, Eq.(2.17) in the text
in (B4)\cite{43}. Thereafter, some simple algebraic manipulations are
done to obtain
\begin{equation} 
\frac{\partial \tilde{\mathcal{N}}} {\partial \Psi\ \ }\ =\ 
\frac{2} {\beta}\  I(0, \Psi, \beta)\ +\ 
\frac{4} {\beta}\ \sum_{s\ =
1}^{\infty} \cos{(2 \pi s x_{0})}\  I(s, \Psi, \beta).   
\end{equation}
\noindent
We note that (B5) is identical to Eq.(2.20) in the text. Furthermore, we
have from the definition that ${\tilde{\mathcal{N}}}(0, \beta, x_{0}) \ =\
0$. See Eq.(2.6) in the text. After integration of (B5) over
$\Psi^{\prime}\ \in (0, \Psi)$ we get Eq.(2.25) in the text. When $\Psi\
=\ \sinh^{2}{n \beta}, \ n\ \in Z$ in Eq.(2.25), we get
${\tilde{\mathcal{N}}}(\Psi, \beta, x_{0}) \ =\ 2 n^{2} \beta$. See
Eq.(3.1) in this context. This particular result has been
obtained by another route in the literature\cite{17, 36}. We consider now
$\nu = 0$ or 
the AL equation. Then from Eqs.(2.13), (2.20), (2.23), (2.24) and
(2.26) we get for $n \in N$
\begin{equation}
\tanh{n \beta} \coth{\beta}\ -\ n \ =\ 0.
\end{equation}
We see then that if $n > 1 $, (B6) has no real nonzero $\beta$ as a
solution.

\section{Application of the method of {\it{successive substitutions}} in
the investigation of the formation of  stationary localized states in
IN-DNLS}

Following the text, we take ${\tilde{\mathcal{N}}}(\Psi, \beta, x_{0})\ =\
2\ \alpha^{2}\ =\ {\rm{Constant}}$. We further write Eq.(2.25) in the form
$\Psi\ =\ F(\Psi)$ where the function $F(\Psi)$ is defined as
\begin{equation}
F(\Psi)\ =\ \sinh^{2}[{\sqrt{\alpha^{2}\ \beta\ -\ f_{2}(\beta, x_{0},
\Psi)}}],
\end{equation}

\noindent
and in (C1)
\[f_{2}(\beta, x_{0}, \Psi)\  =\  2\ \beta \cos{[2 \pi  x_{0}]}\
\frac{\sin^{2}({\frac{\pi} {\beta}}\ {\rm{arc}}\sinh{\sqrt{\Psi}})}
{\sinh{\frac{\pi^{2}} {\beta}}}.\] 
It should be noted that only the first term in the sum in Eq.(2.25) is
retained to obtain (C1). To obtain roots of the equation, $\Psi\ =\
F(\Psi)$, we can use the method of {\it{successive substitutions}}. In
this method at the $k$-th iteration, we write $\Psi_{k + 1}\ =\
F(\Psi_{k})$ with the assumption that $\lim_{k\ \rightarrow\ \infty}
\Psi_{k}\ \rightarrow\ \Psi_{root}$. However, the necessary condition for
this to happen is that $|F^{\prime}(\Psi_{root})|\ <\ 1$\cite{60}. To
explain the use of this method in the  calculation, we shall restrict
ourselves only to the first iteration with $\Psi_{0}\ =\ \sinh^{2}{\alpha
\sqrt{\beta}}$. This in turn implies
\begin{equation}
\Psi\ \sim\ \Psi_{1}\ =\ \sinh^{2}[{\sqrt{\alpha^{2}\ \beta\ -\ 
f_{2}(\beta, x_{0}, \Psi_{0})}}]. 
\end{equation}
\noindent
Note that in the calculation of roots (Eq.(3.11) in the text) we have
taken $f_{2} = 0$.  By approximating $\Psi$ by $\Psi_{1}$ and furthermore 
keeping only the first term in the sum in the definition of $f_{1}(\beta,
\nu, \lambda, x_{0})$ (Eq. (2.21)) we get from Eq.(2.22)
\begin{equation}
{\rm{\tilde{H}}_{0}}\ =\ - 4\ \lambda\ \frac{\Psi_{1}} {\sinh{\beta}}
 - 4\ \lambda\ \nu \frac{\Psi_{1}} {\beta} [1\ +\ 2 \cos{2 \pi x_{0}}\
\frac{\frac{\pi^{2}} {\beta}} {\sinh{\frac{\pi^{2}} {\beta}}}].
\end{equation}}
\noindent
We then get the permissible values of $\beta$ by setting $\frac{d
{\rm{\tilde{H}}_{0}}} {d \beta\ \ }\ =\ 0$.

\end{document}